\documentclass[11pt]{article}
\usepackage{epsfig}
\usepackage{dcolumn}% Align table columns on decimal point
\usepackage{bm}% bold math
\usepackage{graphicx}% Include figure files
\usepackage[textsize=tiny]{todonotes}
\usepackage{subcaption}
\usepackage{here}
\usepackage{amssymb,amsmath}
\usepackage{multirow}
\usepackage{cite,color,url}
\usepackage{tabu}
\usepackage{array}
\usepackage[colorlinks=true,urlcolor=blue,anchorcolor=blue
,citecolor=blue,filecolor=blue,linkcolor=blue,menucolor=blue
,linktocpage=true,pdfproducer=medialab,pdfa=true]{hyperref}
\usepackage{setspace}
\usepackage{colordvi}
\usepackage{epsfig,psfrag,rotating,soul}
\def\beqa{\begin{eqnarray}}
\def\eeqa{\end{eqnarray}}

\evensidemargin \oddsidemargin
\allowdisplaybreaks
\renewcommand{\arraystretch}{1.5}
\psfrag{lk}{$l_k$}
\psfrag{lm}{$l_m$}
\psfrag{blk}{$\bar {l}_k$}
\psfrag{blm}{$\bar {l}_m$}
\def\thefootnote{\fnsymbol{footnote}}
\let\OLDthebibliography\thebibliography
\renewcommand\thebibliography[1]{
\OLDthebibliography{#1}
\setlength{\parskip}{0pt}
\setlength{\itemsep}{0pt plus 0.3ex}}
\usepackage{makecell}

\newcommand{\beq}{\begin{eqnarray}}
\newcommand{\eeq}{\end{eqnarray}}

\newcommand{\GeV}{{\rm\ GeV}}

\def\bsp#1\esp{\begin{split}#1\end{split}}
\def\bpm{\begin{pmatrix}}
	\def\epm{\end{pmatrix}}

\usepackage[normalem]{ulem}

\begin{document}

\thispagestyle{empty}
\begin{center}
\begin{Large}
\textbf{\textsc{Higgs photon associated production  in a Two Higgs Doublet Type-II Seesaw Model at future electron-positron colliders.}}
\end{Large}

\vspace{1cm}
{
B. Ait Ouazghour,$^{1}$%
\footnote{\tt\href{mailto:brahim.aitouazghour@edu.uca.ac.ma}{brahim.aitouazghour@edu.uca.ac.ma}}
M.~Chabab$^{1}$%
\footnote{\tt \href{mailto:mchabab@uca.ac.ma}{mchabab@uca.ac.ma}(corresponding author)}
K.~Goure$^{1}$%
\footnote{\tt \href{mailto:k.goure.ced@uca.ac.ma}{k.goure.ced@uca.ac.ma}}

{\sl $^1$ LPHEA, Faculty of Science Semlalia, Cadi Ayyad University, P.O.B. 2390 Marrakech, Morocco.}}\\

\end{center}

\vspace*{0.1cm}

\begin{abstract}
	We study the one-loop prediction for the single production of a SM-like Higgs boson in association with a photon in electron-positron collisions in the context of the two Higgs doublet type-II seesaw model ($2HDMcT$). We explore to what extent the new scalars in the $2HDMcT$ spectrum affect its production cross-section, the ratio $R_{\gamma h_1}$ as well as the signal strengths $R_{\gamma Z}$ and $R_{\gamma \gamma}$  when $h_1$ is identified with the observed $SM$ Higgs boson within the $2HDMcT$ delimited parameter space. More specifically, we focus on $e^+e^- \to h_1\gamma$ process at one-loop, and analyzed how it evolves under a full set of theoretical constraints and the available experimental data, including $B\to X_s\gamma$ limit at 95$\%$ C.L. Our analysis shows that these observables are strongly dependent on the parameters of the model, especially the mixing angel $\alpha_1$, the potential parameters $\lambda_{7}$, $\lambda_{9}$, the trilinear Higgs couplings, with a noticeable sensitivity to $\alpha_1$. We found that  $\sigma (e^+e^- \to h_1\gamma)$ can  significantly be enhanced  up to $8.1\,\,\times 10^{-2}$ fb, thus exceeding the Standard Model prediction. Additionally, as a byproduct, we also observed that $R_{\gamma h_1}$ is entirely correlated with both the $h_1\to\gamma\gamma$ and $h_1\to\gamma Z$ signal strengths.
 
\end{abstract}
 
\def\thefootnote{\arabic{footnote}}
\setcounter{page}{0}
\setcounter{footnote}{0}
%\newpage
	\section{Introduction}
\label{into}
%%%%%%%
\paragraph*{}
The discovery of the Higgs boson, the last missing piece in the completion of the Standard Model ($SM$),  by the ATLAS \cite{ATLAS:2012yve} and CMS \cite{CMS:2012qbp} collaborations provided an experimental evidence for the Brout-Englert-Higgs mechanism. Since then, major ongoing studies have focused on exploring in detail the properties of the Higgs boson with the aim to enhance our understanding of the fundamental laws of nature. Although most of the $SM$'s predictions have been tested successfully to a high level of accuracy\cite{CMS:2022dwd,ATLAS:2022vkf,Elmetenawee:2024dxc}, it still suffers from drawbacks since it failed to explain several established physical phenomena. As examples, the origin of dark matter\cite{Zwicky:1933gu,Rubin:1970zza}, the hierarchy problem \cite{Veltman:1980mj}, and the neutrino mass generation \cite{RevModPhys.88.030501,RevModPhys.88.030502} do not fit in the $SM$.

\paragraph*{}To address these issues, a plethora of  new physics models have been proposed in literature. Among these beyond the Standard Model ($BSM$), scenarios relying on an extended Higgs sectors with  new scalar fields, as the popular two Higgs Doublet Model ($2HDM$) \cite{Deshpande:1977rw,PhysRevD.15.1958,Branco:2011iw,Dawson:2018dcd,Ivanov:2017dad,PhysRevLett.43.1566}, the Higgs Triplet Models ($HTM$) \cite{PhysRevLett.43.1566,Dawson:2018dcd} and recently 2HDM augmented by a complex triplet scalar, dubbed the two Higgs Doublet Type-II Seesaw Model ($2HDMcT$) \cite{Ouazghour:2018mld,Ouazghour:2023eqr}. Since the spectra of these models generally predict additional Higgs bosons with novel features, searching for these new scalars has been actively conducted as a key focus and one of the major motivations for the current and future experiments, especially at LHC \cite{ATLAS:2017eiz,ATLAS:2017ayi,CMS:2018amk,ATLAS:2018oht,CMS:2019pzc,CMS:2019mij,2022arXiv220701046C}. Besides, as no direct evidence for new physics has been seen yet, precision measurements of the Higgs boson properties  and couplings \cite{Gupta:2013zza,Baglio:2016bop,Tian:2016qlk,Durig:2016jrs,Liu:2018peg} to other new scalars can offer a promising opportunity for a potential discovery of new physics. Indeed, more accurate understanding of the Higgs boson and the measurements of its couplings with high levels of precision is the main goal of future $e^+e^-$ colliders \cite{Moortgat-Pick:2015lbx} such as the International Linear Collider (ILC) \cite{ILCInternationalDevelopmentTeam:2022izu,Bambade:2019fyw}, Compact Linear Collider (CLIC) \cite{CLICPhysicsWorkingGroup:2004qvu,Moortgat-Pick:2015lbx}, Circular Electron-Positron Collider
(CEPC) \cite{CEPCStudyGroup:2018rmc,CEPCStudyGroup:2018ghi,CEPCStudyGroup:2023quu}, and Future Circular Collider (FCC) \cite{TLEPDesignStudyWorkingGroup:2013myl,FCC:2018byv}. Compared to
hadron colliders, these colliders, featuring a cleaner $e^+e^-$ background, can yield substantial improvements over LHC measurements \cite{LCCPhysicsWorkingGroup:2019fvj}.

\paragraph*{}The associated production of the SM-like Higgs boson with a photon, $e^+e^- \to h_1\gamma$ is well suited to study the Higgs-gauge bosons couplings such as the $h_1\gamma\gamma$ and $h_1\gamma Z$ couplings. Since the production rate can be sizably amplified in the presence of new physics contributions as compared to the SM, this process may serve as a potential discovery channel. This process was investigated in the SM \cite{PhysRevD.52.3919,1997NuPhB.491...68D,Barroso:1985et} and recently in some $BSM$ scenarios as the inert Higgs doublet model \cite{Arhrib:2014pva}, $HTM$ \cite{Rahili:2019ixf}, the minimal supersymmetric standard model (MSSM) \cite{2016EPJC,Demirci:2019ush} and the effective field theory \cite{Aoki:2022dxg}.  

\paragraph*{} In this work, we investigate the single production of the neutral Higgs boson in association with a photon in electron positron collisions within the Two Higgs Doublet Type II Seesaw Model ($2HDMcT$). To do that, we implement  a full set of theoretical constraints originated from perturbative unitarity, electroweak vacuum stability, as well as the experimental Higgs exclusion limits from LEP, LHC and Tevatron. Since the  mass generation from seesaw mechanism is similar to the Brout-Englert-Higgs mechanism, 2HDMcT model is appealing, displaying many phenomenological features especially different from those emerging in 2HDM scalar sector. Apart its broader spectrum than 2HDM's one, the doubly charged Higgs $H^{++}$, as a smoking gun of  $2HDMcT$, $H^{++}$ is currently intensively searched for at ATLAS and CMS, by means of promising decay channels, such as $H^{++}H^{--}$ and $H_i^{+}H_i^{-}$ $(i=1,2)$, decaying to the same sign di-lepton \cite{Ouazghour:2018mld}. Furthermore, 2HDMcT is arguably one of the simplest frameworks that can account for both the dark matter issue \cite{Chen:2014lla} as well as the neutrino mass problem \cite{FileviezPerez:2008jbu,Cai:2017mow,King:2015aea}. Along side the doubly charged Higgs the phenomenological features of this model include the possibility of enhanced Higgs couplings, modified Higgs decay channels, and the presence of additional scalar particles that can be probed at the Large Hadron Collider (LHC) or other future colliders. For example one of the prospective signals that can be probed is $pp \to Z/\gamma \to H_2^+ H_2^- \to H^{++} W^- H^{--} W^+ \to l^+l^+l^-l^-+4j$. This process does not show up neither in $2HDM$ nor  in $HTM$ giving that the mass splitting $\Delta m = m_{H^{++}} -  m_{H^+}$ is constrained by the oblique parameters to be less than  $40$ GeV in $HTM$ \cite{Ashanujjaman:2021txz}. New contribution to the oblique parameters from the new states in 2HDMcT makes this avoidable. Additionally, beyond Higgs phenomenology, it has been demonstrated that interactions between doublet and triplet fields may induce a strong first order electroweak phase transition, which provide conditions for the generation of the baryon asymmetry through electroweak baryongenesis \cite{Ramsey-Musolf:2019lsf}. 

%%%%%%%%%%%%%%%%%%%%%%%%%%%%
\paragraph*{}This paper is organized as follows:  In Sect. \ref{section2}, we briefly review $2HDMcT$  model and mention some theoretical and experimental constraints imposed on the model parameter space. In Sect. \ref{section3} we study the two processes $e^+e^- \to h_1\gamma$ and $e^-\gamma  \to h_1 e^-$ in 2HDMcT and discuss the correlation of the signal strengths for the one loop induced processes $h_1 \to \gamma\gamma$ and $h_1 \to  Z\gamma$ in 2HDMcT. Sect. \ref{conlusion} is devoted to our conclusion.

\section{Two Higgs Doublet Type-II Seesaw Model: Brief overview}
\label{section2}
 2HDMcT has recently gained interest as a well-motivated extension of the 2HDM. Besides the two Higgs doublets with hypercharge $Y=+1$, 2HDM is augmented with one colorless triplet field $\Delta$ transforming under the $SU(2)_L$ gauge group as a complex scalar with $Y_\Delta=2$. 
\begin{equation}
\Phi_1 = \left(
\begin{array}{c}
\phi_1^+ \\
\phi_1^0 \\
\end{array}
\right)
\quad {\rm ,}\quad 
\Phi_2 = \left(
\begin{array}{c}
\phi_2^+ \\
\phi_2^0 \\
\end{array}
\right)
\quad {\rm and}\quad 
\begin{matrix}
\Delta &=&\left(
\begin{array}{cc}
\delta^+/\sqrt{2} & \delta^{++} \\
(v_t+\delta^0+i\eta_0)/\sqrt{2} & -\delta^+/\sqrt{2}\\
\end{array}
\right)
\end{matrix}
\end{equation}
with $\phi_1^0 = (v_1+\psi_1+ i \eta_1)/\sqrt{2}$, $\phi_2^0 = (v_2+\psi_2+ i \eta_2)/\sqrt{2}$. $v_1$, $v_2$ and $v_t$ denote  the vacuum expectation values of the Higgs doublets and triplet fields  respectively, acquired when the  electroweak symmetry is spontaneously broken, with $\sqrt{v_1^2+v_2^2+2v_t^2}=246$ GeV. Consequently, eleven physical Higgs states occur in the model spectrum: three CP-even neutral Higgs bosons $(h_1, h_2, h_3)$, four simply charged Higgs bosons $(H_1^{\pm}, H_2^{\pm})$, two CP odd Higgs bosons $(A_1, A_2)$, and finally two doubly charged Higgs bosons $H^{\pm\pm}$. For details see \cite{Ouazghour:2018mld}.

The most general $SU(2)_L\times U(1)_Y$ invariant scalar potential in this model reads :  \cite{Branco:2011iw,Ouazghour:2018mld}:
\begin{eqnarray}
V(\Phi_1,\Phi_2,\Delta) &=& m_{11}^2 \Phi_1^\dagger\Phi_1+m_{22}^2\Phi_2^\dagger\Phi_2-[m_{12}^2\Phi_1^\dagger\Phi_2+{\rm h.c.}]+\frac{\lambda_1}{2}(\Phi_1^\dagger\Phi_1)^2
+\frac{\lambda_2}{2}(\Phi_2^\dagger\Phi_2)^2
\nonumber\\
&+& \lambda_4(\Phi_1^\dagger\Phi_2)(\Phi_2^\dagger\Phi_1)+ \left\{\frac{\lambda_5}{2}(\Phi_1^\dagger\Phi_2)^2
+\big[\beta_1(\Phi_1^\dagger\Phi_1)
+\beta_2(\Phi_2^\dagger\Phi_2)\big]
\Phi_1^\dagger\Phi_2+{\rm h.c.}\right\} \nonumber\\
&+&\lambda_3(\Phi_1^\dagger\Phi_1)(\Phi_2^\dagger\Phi_2)+\lambda_6\,\Phi_1^\dagger \Phi_1 Tr\Delta^{\dagger}{\Delta} +\lambda_7\,\Phi_2^\dagger \Phi_2 Tr\Delta^{\dagger}{\Delta}\nonumber\\
&+&\left\{\mu_1 \Phi_1^T{i}\sigma^2\Delta^{\dagger}\Phi_1 + \mu_2\Phi_2^T{i}\sigma^2\Delta^{\dagger}\Phi_2 + \mu_3 \Phi_1^T{i}\sigma^2\Delta^{\dagger}\Phi_2  + {\rm h.c.}\right\}+\lambda_8\,\Phi_1^\dagger{\Delta}\Delta^{\dagger} \Phi_1\nonumber\\
&+&\lambda_9\,\Phi_2^\dagger{\Delta}\Delta^{\dagger} \Phi_2+m^2_{\Delta}\, Tr(\Delta^{\dagger}{\Delta}) +\bar{\lambda}_8(Tr\Delta^{\dagger}{\Delta})^2\hspace*{0cm}+\hspace*{0cm}\bar{\lambda}_9Tr(\Delta^{\dagger}{\Delta})^2
\label{scalar_pot}
\end{eqnarray}
In this work we assume that $m_{11}^2$, $m_{22}^2$, $m_{\Delta}^2$, $m_{12}^2$, $\lambda_{1,2,3,4,5,6,7,8,9}$, $\bar{\lambda}_{8,9}$, $\mu_{1,2,3}$, $\beta_{1,2}$ are real parameters. To avoid tree-level Higgs mediated $FCNC_s$ at tree level, we consider $Z_2$ symmetry where $\beta_1=\beta_2=0$. Also the $Z_2$ symmetry is  softly broken by the bi-linear terms proportional to $m_{12}^2$, $\mu_1$, $\mu_2$ and $\mu_3$ parameters. Thanks to the combination $v^2=v_1^2+v_2^2+2 v_t^2=(2\sqrt{2} G_F)^{-1}$ and the three minimization conditions, the scalar potential Eq. (\ref{scalar_pot}) has seventeen independent parameters, one possible choice is :
\begin{center}
	$\alpha_1$, $\alpha_2$, $\alpha_3$, $m_{h_1}$, $m_{h_2}$, $m_{h_3}$, $m_{H^{\pm\pm}}$, $ \lambda_{1}$, $\lambda_{3}$, $\lambda_{4}$, $\lambda_{6}$, $\lambda_{8}$, $\bar{\lambda}_{8}$, $\bar{\lambda}_{9}$, $\mu_1$, $v_t$ and $\tan\beta$ 
\end{center}
where $\alpha_{i=1,2,3}$ are the CP-even mixing angles and $\tan\beta=v_2/v_1$.
\paragraph*{} In our subsequent analysis $h_1$ is identified to the SM-like Higgs boson with $m_{H^{SM}}=125$ GeV \cite{ATLAS:2012yve,CMS:2012qbp}. So the scalar potential is described by sixteen free parameters. The Yukawa lagrangian $\mathcal{L}_{\rm Yukawa}$ contains all the Yukawa sector of the Two Higgs Doublet Model plus one extra Yukawa term emerging from the triplet field and generating  a small Majorana mass terms for the neutrinos when the symmetry is spontaneously  broken. 
\begin{equation}
-\mathcal{L}_{\rm Yukawa} \supset  - Y_{\nu} L^T C \otimes i \sigma^2 \Delta L  + {\rm h.c.} \label{eq:yukawa}
\end{equation}
For the type II 2HDMcT, up quarks interact
with $\Phi_2$, while leptons and down quarks with $\Phi_1$ as :
\begin{eqnarray}
-{\cal L}^{II}_{Yukawa} &=& -y_{u} \bar{Q}_{L} \tilde{\Phi}_{2} u_{R}-y_{d} \bar{Q}_{L} \Phi_{1}d_{R}-y_{\ell} \bar{L}_{L} \Phi_{1} \ell_{R}+h.c\,, \label{eq:Yukawa_CH}
\end{eqnarray} 
Here $Q_{L}$ and $L_{L}$ represent the left-handed quark and lepton doublets, $u_{R}$, $d_{R}$, and $\ell_{R}$ are the right-handed up-type quarks, down-type quarks, and lepton singlets. $\Phi_{1}$ and $\Phi_{2}$ are the two Higgs doublets , with $\tilde{\Phi}_{2}=i \sigma_{2} \Phi_{2}^{*}$. $y_{u}$, $y_{d}$, and $y_{\ell}$ are the Yukawa couplings for the up, down-type quarks and leptons, respectively. The Yukawa couplings of the CP-even Higgs bosons $h_i$ to fermions are presented in Table \ref{table1}. The matrix elements $\mathcal{E}_{ij}$ are given in Appendix \ref{scalar}. 
\begin{table}[H]
	\begin{center}
		\setlength{\tabcolsep}{8pt}
		\begin{tabular}{c|c|c|c} \hline\hline 
			$\phi$  & $\xi^u_{\phi}$ &  $\xi^d_{\phi}$ &  $\xi^\ell_{\phi}$  \\   \hline
			$h_1$~ 
			& ~ $  \frac{\mathcal{E}_{12}}{s_\beta}$~
			& ~ $ -\frac{\mathcal{E}_{11}}{c_\beta} $~
			& ~ $ -\frac{\mathcal{E}_{11}}{c_\beta} $ ~ \\
			$h_2$~
			& ~ $  \frac{\mathcal{E}_{22}}{s_\beta}$~
			& ~ $ -\frac{\mathcal{E}_{21}}{c_\beta} $~
			& ~ $ -\frac{\mathcal{E}_{21}}{c_\beta} $ ~ \\
			$h_3$~
			& ~ $  \frac{\mathcal{E}_{32}}{s_\beta}$~
			& ~ $ -\frac{\mathcal{E}_{31}}{c_\beta} $~
			& ~ $ -\frac{\mathcal{E}_{31}}{c_\beta} $ ~ \\ \hline \hline 
		\end{tabular}
	\end{center}
	\caption {Yukawa couplings of the $h_1$, $h_2$, and $h_3$ bosons to the quarks and leptons in 2HDMcT} 
	\label{table1}
\end{table} 
\subsection{Theoretical and Experimental Constraints}
\label{constraint}
\paragraph*{}
The phenomenological analysis in $2HDMcT$ is performed via  implementation of a full set of theoretical constraints ~\cite{Ouazghour:2018mld, Ouazghour:2023eqr} as well as the Higgs exclusion limits from various experimental measurements at colliders, namely :
\begin{itemize}
	\item \textbf{Unitarity} : The scattering processes must obey  perturbative unitarity.
	\item \textbf{Perturbativity}: The quartic couplings of the scalar potential are constrained by the following conditions : $| \lambda_i|<8 \pi$ for each $i=1,..,5$.
	\item \textbf{Vacuum stability} : Boundedness from below $BFB$ arising from the positivity in any direction of the fields $\Phi_i$, $\Delta$.
     \item[\textbullet]{\bf Electroweak precision observables}: The oblique parameters $S, T$ and $U$~\cite{Peskin:1991sw,Grimus:2008nb} have been calculated in $2HDMcT$~\cite{Ouazghour:2023eqr}.  The analysis of the precision electroweak data in light of the new PDG mass of the $W$ boson yields~\cite{ParticleDataGroup:2022pth}:	
	\begin{eqnarray}
		\widehat S_0= -0.01\pm 0.07,\  \ \widehat T_0 = 0.04\pm 0.06,\ \ \rho_{ST} = 0.92,  \nonumber 
	\end{eqnarray}	
We use the following $\chi^2_{ST}$ test :

\begin{equation}
	\label{eq:STRange}
	\frac{(S-\widehat S_0)^2}{\sigma_S^2}\ +\
	\frac{(T-\widehat T_0)^2}{\sigma_T^2}\ -\
	2\rho_{ST}\frac{(S-\widehat S_0)(T-\widehat T_0)}{\sigma_S \sigma_T}\
	\leq\ R^2\,(1-\rho_{ST}^2)\; ,
\end{equation}
with $R^2=2.3$ and $5.99$ corresponding to $68.3 \%$  and
$95 \% $  confidence levels (C.L.) respectively.
 Our numerical analysis is performed with $\chi^2_{ST}$ at 95\% C.L. 
    \item To further delimit the allowed parameter space, the \texttt{HiggsTools} package~\cite{Bahl:2022igd} is employed. This ensures that the allowed parameter regions align with the observed properties of the $125$~GeV Higgs boson  ( \texttt{HiggsSignals}~\cite{Bechtle:2013xfa,Bechtle:2014ewa,Bechtle:2020uwn,Bahl:2022igd}) and with the limits from searches for additional Higgs bosons at the LHC and at LEP ( \texttt{HiggsBounds}~\cite{Bechtle:2008jh,Bechtle:2011sb,Bechtle:2013wla,Bechtle:2020pkv,Bahl:2022igd}).
    \item[\textbullet]{\bf Flavour constraints}:  Flavour constraints are also implemented in our analysis. We used $B$-physics results, derived in \cite{Ouazghour:2023eqr} as well as  the experimental data at 2$\sigma$ \cite{HFLAV:2022pwe} displayed in Table \ref{Tab2}.
\end{itemize}
{\renewcommand{\arraystretch}{1.5} %donne la distance entre les lignes%
	{\setlength{\tabcolsep}{0.1cm} %donne la distance entre les collones%
		\begin{table}[H]
			\centering
			\setlength{\tabcolsep}{7pt}
			\renewcommand{\arraystretch}{1.2} %
			\begin{tabular}{|l||c|c|}
				\hline
				Observable & Experimental result & 95\% C.L.\\\hline
				BR($\bar{B}\to X_{s}\gamma$)\cite{Ouazghour:2023eqr}&$(3.49\pm 0.19)\times10^{-4}$\cite{HFLAV:2022pwe}&$[3.11\times 10^{-4} , 3.87\times 10^{-4}]$\\\hline
			\end{tabular}
			\caption{Experimental result of flavor observable: $\bar{B}\to X_{s}\gamma$ at 95$\%$ C.L.}
			\label{Tab2}
		\end{table}
	%	\hspace{1cm}
	
		\section{$e^+e^-\to \gamma h_1$ and $e^-\gamma\to e^-h_1$ in 2HDMcT}
		\label{section3}
		\subsection{Process  topology}
		The $e^+e^-\to \gamma h_1$ process has been studied in many of beyond the Standard Model frameworks. In the $2HDMcT$, at tree level, the associated production processes $e^+e^- \to \gamma h_1$ and $e^- \gamma \to e^- h_1$ are intermediated by the t-channel and s-channel electron exchange diagrams respectively. However, the former diagrams are suppressed by the electron mass. At one-loop these diagrams are mediated by the self-energy, box, and triangle diagrams and hence  are sensitive to all virtual particles inside the loops including the charged Higgs states, $H^\pm_1$, $H^\pm_2$ and $H^{\pm\pm}$, predicted by our model. 
		%------------------------------------------------------------------------------------
		\begin{figure}[H]	
			%	\begin{minipage}{0.51\textwidth}
			\centering
			\includegraphics[scale=0.46]{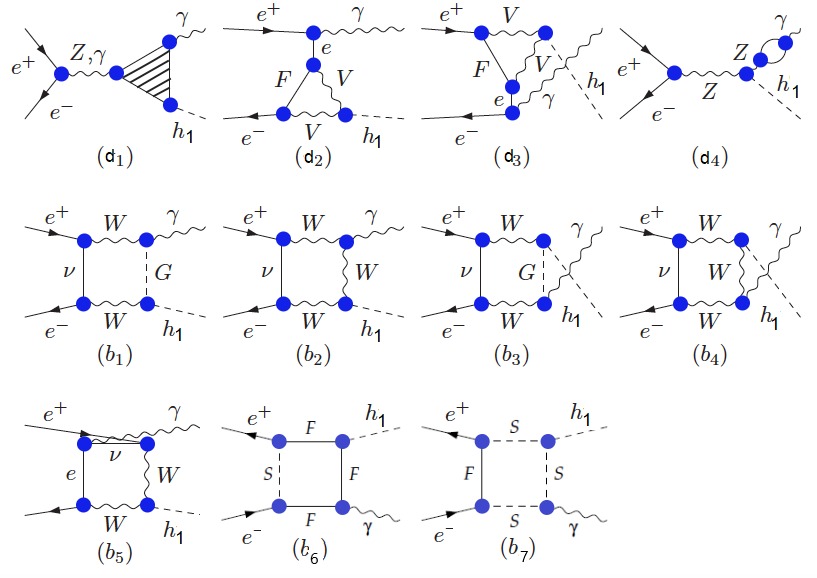}
			%	\end{minipage}	
			\caption{Feynman diagrams involving the charged contributions to $e^-\,e^+\,\,\to\,\,\gamma\, h_1$ process in \textsc{2HDMcT}. In $d_{1}$, $d_{4}$, $b_{6}$ and $b_{7}$ diagrams, the loops receive contributions from SM particles as well as the charged Higgs bosons $H^\pm_1$, $H^\pm_2$ and $H^{\pm\pm}$. Generic diagrams $d_{2},d_{3}$, depict $h_1V V$, $V=\gamma,\;Z$.}
			\label{fig3}
		\end{figure}
		\begin{figure}[H]	
		%	\begin{minipage}{0.51\textwidth}
		\centering
		\includegraphics[scale=0.29]{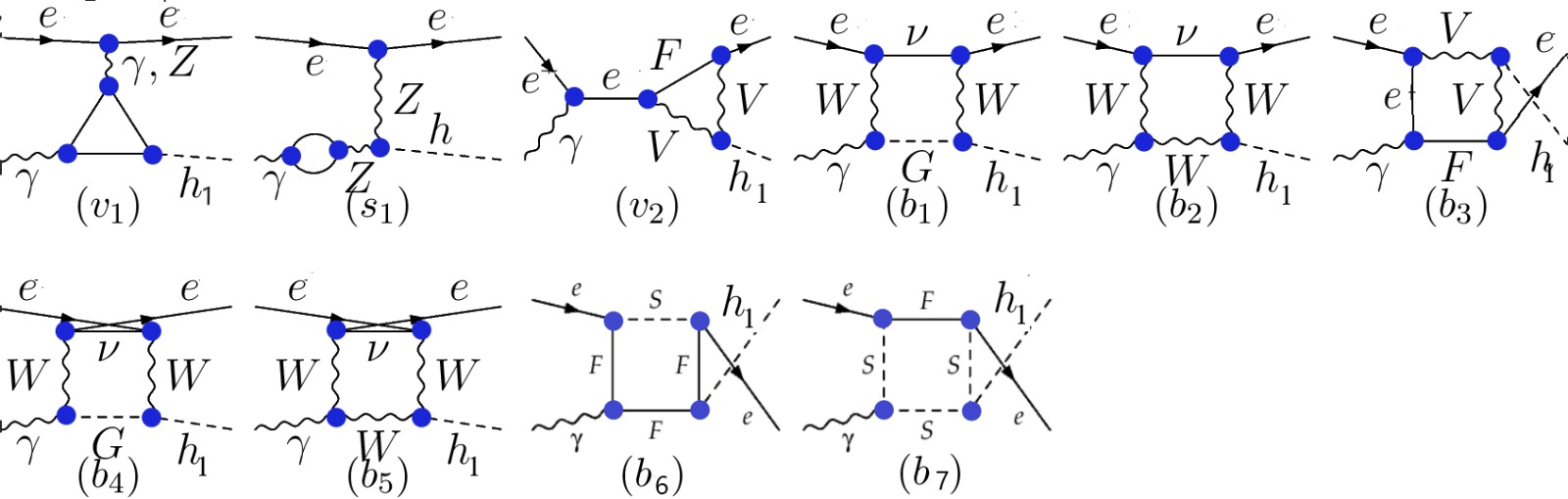}
		%	\end{minipage}	
		\caption{Feynman diagrams involving the charged contributions to $e^- \gamma \to e^- h_1$ process in the \textsc{2HDMcT}. In diagrams $v_{1}$, we depict all possible charged particles. Diagrams $s_1$, $b_6$ and $b_7$ depict the contributions from  $H^\pm_1$, $H^\pm_2$ and $H^{\pm\pm}$ bosons.}
		\label{fig44}
	\end{figure}
		%---------------------------------------------------------
		Figs. \ref{fig3}, \ref{fig44} and \ref{fig4} illustrate the generic Feynman diagrams that effectively contribute to the $e^+e^- \to \gamma h_1$, $e^- \gamma \to e^- h_1$ and  $h_1 \to \gamma V$  processes, with $V=\gamma,\;Z$. Here $F$ stands for any fermionic particle, while $S$ denotes the charged states $H^\pm_1$, $H^\pm_2$ and $H^{\pm\pm}$.
		\begin{figure}[H]	
			%	\begin{minipage}{0.51\textwidth}
			\centering
			\includegraphics[scale=0.53]{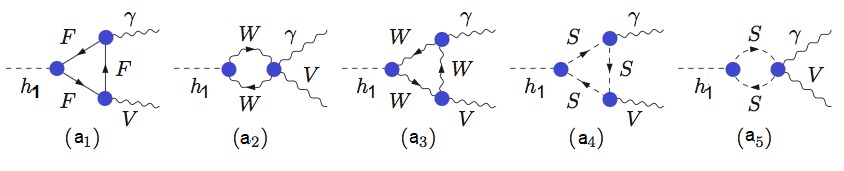}
			%	\end{minipage}	
			\caption{ Feynman diagrams for the decay process $h_1 \to \gamma V$ ($V=\gamma,\;Z$) in the $2HDMcT$.  $a_{4},a_{5}$ illustrate the contributions of the charged Higgses $H^\pm_1$, $H^\pm_2$ and $H^{\pm\pm}$.}
			\label{fig4}
		\end{figure}
    
    Summing over the polarization of the photon, the differential cross section of the Higgs associated production with a photon is  calculated using the the generic formula :
		\begin{equation} \label{cross1}
			\frac{d\sigma}{d\cos\theta}(e^+ e^- \rightarrow h_1 \gamma)=\frac
			{s-m_{h_1}^2}{32\pi s^{2}} \sum_{pol} |{\cal M}|^2,
		\end{equation}
		where $\sqrt{s}$ is the center-of-mass energy of $e^+e^-$ collisions and $\theta$ is the scattering angle between the photon and the electron in the centre-of-mass frame.
		
		The total amplitudes for the process can be expressed as a sum over all contributions from triangle, box, and self-energy diagrams.
		\begin{equation}\label{eq:totalM}
			{\cal M}= {\cal M}_{\triangle} + {\cal M}_{\Box}+ {\cal M}_{\bigcirc}
		\end{equation}
			\begin{itemize}
		\item ${\cal M}_{\triangle}$ : Contains s-channel $\gamma$ and Z exchange vertex diagrams that involve virtual W and heavy fermion loops. The significant contributions from these fermion loops arise mainly from top and bottom quark loops, as the Yukawa couplings to lighter fermions are very small. Also t–channel vertex diagrams involving W/neutrino and Z/electron exchanges (corrections to the $h_1e^+e^-$ vertex).
		\item  ${\cal M}_{\Box}$ : The box diagrams get contribution from W/neutrino and Z/electron diagrams.
		\item  ${\cal M}_{\bigcirc}$ : Include SM loops contributions from charged fermions and W bosons.
		\item Additionally ${\cal M}_{\triangle,\Box,\bigcirc}$ also receive contributions from the new charged scalars, namely, $H^\pm_1$, $H^\pm_2$ and $H^{\pm\pm}$ as illustrated in Fig. \ref{fig3}. 
	\end{itemize}
	
		 The integrated cross section over all $\theta$ angles is given by
		\begin{equation}  \label{eq:total_cross}
			\sigma(e^+ e^- \rightarrow h_1 \gamma)=\int_{-1}^{+1} d\cos\theta \frac{d\sigma}{d\cos\theta}.
		\end{equation}
	
		Our calculation is performed using dimensional regularization 
		along with {\sl FeynArts} and {\sl FormCalc} packages \cite{Hahn:2000kx,HAHN1999153}. 
		while the numerical evaluation of the scalar integrals is done with 
		{\sl LoopTools} \cite{vanOldenborgh:1990yc,Hahn:2010zi}. Note that the gauge invariance of our final results is assured once contributions from all box and triangle diagrams are summed.  In the subsequent analysis, we introduce the ratio,
		\begin{eqnarray}
		R_{\gamma h_1}  \equiv  \frac{\sigma(e^+e^-\to \gamma h_1)}{\sigma_{\rm}(e^+e^- \to \gamma H^{\rm SM})}
		\end{eqnarray}
		which is the total cross-section in the $2HDMcT$ normalized to the $SM$ one. Also we define the diphoton and $\gamma Z$ gauge bosons signal strengths, $R_{\gamma\gamma}$ and $R_{Z\gamma}$,
		\begin{equation}
		R_{\gamma \gamma(Z\gamma)}  \equiv  \frac{\sigma(gg \to h_1)\times Br(h_1\to\gamma \gamma (Z\gamma))}{\sigma(gg \to H^{\rm SM})\times Br(H^{\rm SM} \to\gamma \gamma(Z\gamma))}
		\end{equation}
	    The decay width for $h_1\to\gamma\gamma$ and $h_1\to\gamma Z$ in 2HDMcT are given by,
		\begin{eqnarray}
			&& \Gamma(h_1 \rightarrow\gamma\gamma) = \frac{G_F\alpha^2 M_{{h_1}}^3}
			{128\sqrt{2}\pi^3} \bigg| \sum_f N_c Q_f^2 k_f 
			A_{1/2}^{{h_1}} (\tau_f) + \lambda_{h_1WW} A_1^{{h_1}} (\tau_W)+\tilde{\lambda}_{h_1 H_1^\pm\,H_1^\mp}
			A_0^{{h_1}}(\tau_{H_1^{\pm}}) + \tilde{\lambda}_{h_1 H_2^\pm\,H_2^\mp}
			A_0^{{h_1}}(\tau_{H_2^{\pm}}) \nonumber \\
			&&\hspace{1.9cm}\,+\, 4 \tilde{\lambda}_{h_1 H^{\pm\pm}H^{\mp\mp}}
			A_0^{{h_1}}(\tau_{H^{\pm\pm}}) \bigg|^2
			\label{partial_width_htm}\\
			&& \Gamma (h_1 \to Z\gamma) = \frac{G^2_\mu M_W^2\,\alpha\,M_{h_1}^{3}}
			{64\,\pi^{4}} \left(1-\frac{M_Z^2}{M_{h_1}^2} \right)^3 \bigg| \sum_{f} 
			\frac{Q_f\, \hat{v}_f N_c}{c_W} \,k_f\, {\cal F}^{h_1}_{1/2} 
			(\tau_f,\lambda_f) + \lambda_{WW}\,{\cal F}^{h_1}_1 (\tau_W,\lambda_W) \nonumber\\
			&&\hspace{1.9cm}\,+\, \lambda_{ZH_1^{\pm}H_1^{\mp}}\tilde{\lambda}_{{h_1} H_1^\pm\,H_1^\mp}\,{\cal F}^{h_1}_0 (\tau_{H_1^{\pm}},\lambda_{H_1^{\pm}}) + \lambda_{ZH_2^{\pm}H_2^{\mp}}\tilde{\lambda}_{{h_1} H_2^\pm\,H_2^\mp}\,{\cal F}^{h_1}_0 (\tau_{H_2^{\pm}},\lambda_{H_2^{\pm}})\nonumber\\ 
			&&\hspace{1.9cm}\,+\, \lambda_{ZH^{\pm\pm}H^{\mp\mp}}\tilde{\lambda}_{{h_1} H^{\pm\pm}\,H^{\mp\mp}}\,{\cal F}^{h_1}_0 (\tau_{H^{\pm\pm}},
			\lambda_{H^{\pm\pm}}) \bigg|^2 \\
			%&& \Gamma({h_1} \rightarrow gg) = \frac{G_F\alpha^2 M_{{h_1}}^3}{36\sqrt{2}\pi^3} \bigg| \sum_f N_c Q_f^2 k_f 
			%A_{1/2}^{{h_1}}(\tau_f)\bigg|^2
		\end{eqnarray}
	where $\tilde{\lambda}_{h_1 H_i^\pm\,H_i^\mp}=\lambda_{h_1 H_i^\pm\,H_i^\mp}\times\frac{M_W^2}{M_{H_i^{\pm}}^2}$, i=1,2 and $\tilde{\lambda}_{h_1 H^{\pm\pm}H^{\mp\mp}}= {\lambda}_{h_1 H^{\pm\pm}H^{\mp\mp}}\times\frac{M_W^2}{M_{H^{\pm\pm}}^2}$. The functions $A_{1/2}^{h_1}(\tau_i)$, $A_{1}^{h_1}(\tau_i)$, $A_{0}^{h_1}(\tau_i)$, ${\cal F}^{h_1}_0 (\tau_{H^{i}},\lambda_{H^{i}})$, ${\cal F}^{h_1}_1 (\tau_{W},\lambda_{W})$ and ${\cal F}^{h_1}_1 (\tau_{f},\lambda_{f})$ are defined in Appendix \ref{Appendix_A}. The reduced couplings $\lambda_{h_1 ff}$ and $\lambda_{h_1 WW}$ of the Higgs bosons to fermions and $W$ bosons are dispayed in Tables \ref{table1} and \ref{table2} respectively. $\lambda_{ZH_1^{\pm}H_1^{\mp}}$, $\lambda_{ZH_2^{\pm}H_2^{\mp}}$ and $\lambda_{ZH^{\pm\pm}H^{\mp\mp}}$ are given in Eqs. \ref{eq:couplagesZHpZHpp}
\begin{table}[!h]
	\begin{center}
		\setcellgapes{4pt}
		\begin{tabular}{|c|c|c|}
			\hline  
			& $C^{h_i}_W$    & $C^{h_i}_Z$  \\
			\hline  $h_1$ & $\displaystyle{\frac{v_1}{v} \mathcal{E}_{11} + \frac{v_2}{v} \mathcal{E}_{21} + 2\,\frac{v_t}{v} \mathcal{E}_{31}}$ & 
			$\displaystyle{\frac{v_1}{v} \mathcal{E}_{11} + \frac{v_2}{v} \mathcal{E}_{21} + 4\,\frac{v_t}{v} \mathcal{E}_{31}}$  \\
			\hline  $h_2$ & $\displaystyle{\frac{v_1}{v} \mathcal{E}_{12} + \frac{v_2}{v} \mathcal{E}_{22} + 2\,\frac{v_t}{v} \mathcal{E}_{32}}$ & 
			$\displaystyle{\frac{v_1}{v} \mathcal{E}_{12} + \frac{v_2}{v} \mathcal{E}_{22} + 4\,\frac{v_t}{v} \mathcal{E}_{32}}$  \\
			\hline  $h_3$ & $\displaystyle{\frac{v_1}{v} \mathcal{E}_{13} + \frac{v_2}{v} \mathcal{E}_{23} + 2\,\frac{v_t}{v} \mathcal{E}_{33}}$ & 
			$\displaystyle{\frac{v_1}{v} \mathcal{E}_{13} + \frac{v_2}{v} \mathcal{E}_{23} + 4\,\frac{v_t}{v} \mathcal{E}_{33}}$  \\
			\hline 
		\end{tabular}
		\caption{The normalized couplings of the neutral ${\mathcal{CP}}_{even}$ $h_i$ Higgs to the massive gauge bosons $V=W,Z$ in $2HDMcT$. The  matrix elements $\displaystyle{\mathcal{E}_{ij}}$ are given in Eq. \ref{cpevenmatrix}}
		\label{table2}
	\end{center}
\end{table}
}
Compared with the Standard Model, the amplitudes for the loop processes  $e^-\gamma\to e^-h_1$, $e^+e^-\to \gamma h_1$ and $h_1 \to \gamma V$ ($V=\gamma,\;Z$) receive additional contributions from the charged Higgs bosons predicted in the model spectrum $H^\pm_1$, $H^\pm_2$ and $H^{\pm\pm}$. These amplitudes are essentially proportional to the couplings  with $h_1$ Higgs boson, 
			\begin{eqnarray}
			%	\[
			%\begin{matrix}
			\bar{\lambda}_{h_1H^\pm_1H^\pm_1}&=&\frac{s_w}{2e m_w}\left(2 \mathcal{C}_{21}^2 \left(\lambda _6 \mathcal{E}_{13} v_{\Delta }+\lambda _1 v_1 \mathcal{E}_{11}+\lambda _3 v_2 \mathcal{E}_{12}\right)\right.\nonumber\\
			&&+2 \mathcal{C}_{22}^2 \left(\lambda _7 \mathcal{E}_{13} v_{\Delta }+\lambda _2 v_2 \mathcal{E}_{12}+\lambda _3 v_1 \mathcal{E}_{11}\right)\nonumber\\
			&&+\mathcal{C}_{23}^2 \left(2 \mathcal{E}_{13} (2\overline{\lambda _{8}}+\overline{\lambda _{9}}) v_{\Delta }+\left(2 \lambda _6+\lambda _8\right) v_1 \mathcal{E}_{11}+\left(2 \lambda _7+\lambda _9\right) v_2 \mathcal{E}_{12}\right)\nonumber\\
			&&+\mathcal{C}_{22} \mathcal{C}_{23} \left(\sqrt{2} \lambda _9 \mathcal{E}_{12} v_{\Delta }+\sqrt{2} \lambda _9 v_2 \mathcal{E}_{13}-4 \mu _2 \mathcal{E}_{12}-2 \mu _3 \mathcal{E}_{11}\right)\nonumber\\
			&&+\mathcal{C}_{21} \left(\right.\mathcal{C}_{23} \left(\sqrt{2} \lambda _8 \mathcal{E}_{11} v_{\Delta }+\sqrt{2} \lambda _8 v_1 \mathcal{E}_{13}-4 \mu _1 \mathcal{E}_{11}-2 \mu _3 \mathcal{E}_{12}\right)\nonumber\\
			&&\left.+2 \mathcal{C}_{22} \left(\lambda _4+\lambda _5\right) \left(v_2 \mathcal{E}_{11}+v_1 \mathcal{E}_{12}\right)\left.\right)\right)
			%%%%%%%%%%%%%%%%%%%%%%%
			\label{3.3}
		\end{eqnarray}
		\begin{eqnarray}
			\bar{\lambda}_{h_1H^\pm_2H^\pm_2}&=&\frac{s_w}{2e m_w}\left(2 \mathcal{C}_{31}^2 \left(\lambda _6 \mathcal{E}_{13} v_{\Delta }+\lambda _1 v_1 \mathcal{E}_{11}+\lambda _3 v_2 \mathcal{E}_{12}\right)\right.\nonumber\\
			&&+2 \mathcal{C}_{32}^2 \left(\lambda _7 \mathcal{E}_{13} v_{\Delta }+\lambda _2 v_2 \mathcal{E}_{12}+\lambda _3 v_1 \mathcal{E}_{11}\right)\nonumber\\
			&&+\mathcal{C}_{33}^2 \left(2 \mathcal{E}_{13} (2\overline{\lambda _{8}}+\overline{\lambda _{9}}) v_{\Delta }+\left(2 \lambda _6+\lambda _8\right) v_1 \mathcal{E}_{11}+\left(2 \lambda _7+\lambda _9\right) v_2 \mathcal{E}_{12}\right)\nonumber\\
			&&+\mathcal{C}_{32} \mathcal{C}_{33} \left(\sqrt{2} \lambda _9 \mathcal{E}_{12} v_{\Delta }+\sqrt{2} \lambda _9 v_2 \mathcal{E}_{13}-4 \mu _2 \mathcal{E}_{12}-2 \mu _3 \mathcal{E}_{11}\right)\nonumber\\
			&&+\mathcal{C}_{31} \left(\right.\mathcal{C}_{33} \left(\sqrt{2} \lambda _8 \mathcal{E}_{11} v_{\Delta }+\sqrt{2} \lambda _8 v_1 \mathcal{E}_{13}-4 \mu _1 \mathcal{E}_{11}-2 \mu _3 \mathcal{E}_{12}\right)\nonumber\\
			&&\left.+2 \mathcal{C}_{32} \left(\lambda _4+\lambda _5\right) \left(v_2 \mathcal{E}_{11}+v_1 \mathcal{E}_{12}\right)\left.\right)\right)
			\label{3.4}
		\end{eqnarray}
		\begin{eqnarray}
			\bar{\lambda}_{h_1H^{\pm\pm}H^{\pm\pm}}&=&\frac{s_w}{e m_w}\left(2 \overline{\lambda _{8}} \mathcal{E}_{13} v_{\Delta }+\lambda _6 v_1 \mathcal{E}_{11}+\lambda _7 v_2 \mathcal{E}_{12}\right)
			\label{3.5}
		\end{eqnarray}
		At this stage it is worth noticing that:
			\begin{itemize}
			\item  These couplings are mostly fixed by  $\alpha_1$, $\lambda_{3}$, $\lambda_{4}$, $\lambda_7$ and $\lambda_9$ parameters when $\tan\beta$ takes relatively large values, since $v_t<<v_1<v_2$.
			 \item Depending on their signs, the charged Higgs contributions can either enhance or suppress the $e^+e^-\to \gamma h_1$ and $h_1 \to \gamma V$ rates ($V=\gamma,\;Z$).
		\end{itemize}
	%	\newpage
		\subsection{Results and analysis}
		\paragraph*{} Given that $h_1$ is identified as the SM-like Higgs boson, observed at the LHC with $m_{h_1}=125$ GeV, we perform scans within the allowed parameter space  by implementation of the theoretical and experimental constraints mentioned above.\\
		The following set of input parameters  is used in the subsequent numerical analysis :
		\begin{eqnarray}
			\mathcal{P}_I = \left\{\alpha_1,\alpha_2,\alpha_3,m_{h_1},m_{h_2}, \lambda_{1},\lambda_{3},\lambda_{4},\lambda_{6},\lambda_{7},\lambda_{8},\lambda_{9},\bar{\lambda}_{8},\bar{\lambda}_{9},\mu_1,v_t,\tan\beta \right\}
			\label{eq:set-para1} 
		\end{eqnarray}
		\begin{eqnarray}
		\begin{matrix}
		     m_{h_1}=125 \,\,\text{GeV}, \,\,\,  m_{h_1}\leq m_{h_2}\leq m_{h_3}\leq 1\,\text{TeV},\,\,\,

	      	80\,\,\text{GeV}\leq m_{H^\pm_1}, m_{H^\pm_2},  m_{H^{\pm\pm}}\leq 1\,\text{TeV},\,\, \\

	      	  -2\pi \leq\lambda_1\leq 2\pi,\,\,-2\pi \leq \lambda_{3,4}\leq 2\pi , \,\, \,\,\,-2\pi\leq \lambda_{6, 7}\leq 2\pi,\,\,\,-2\pi\leq \lambda_{8,9}\leq 2\pi, \,\, -2\pi\leq \bar{\lambda}_{8,9} \leq 2\pi \\
	      		-\pi/2 \leq \alpha_1 \leq  \pi/2, \,\, \alpha_2\approx 0, \,\, \, \alpha_3\approx 0, \,\,\mu_1= v_t = 1 \,\,\text{GeV}, \,\, \tan\beta= 8
	  \end{matrix}
	      		\label{11}
	\end{eqnarray}
		Since $v_t<<v_1<v_2$,  we can see from equations (\ref{3.3}), (\ref{3.4}), and (\ref{3.5}) , that the virtual charged states $H_1^\pm $, $H_2^\pm$ impact on $h_1\gamma\gamma$ and $h_1\gamma Z$ couplings induce a dependence on $\lambda_{3}$, $\lambda_{4}$, $\lambda_7$ and $\lambda_9$, with a noticeable  sensitivity to $\alpha_1$. The doubly charged  Higgs $H^{\pm\pm}$ instead leads to $\lambda_{7}$ dependence and to $\alpha_1$ strong sensitivity.  Comparatively, the influence of the  other model parameters is either mild or negligible.
		\paragraph*{}
		In Fig. \ref{corre_betw_coupla}, we show the allowed ranges for $(\lambda_7,\,\,\lambda_9)$ as well as the size of the trilinear couplings $\bar{\lambda}_{h_1 H_1^\pm H_1^\pm}$, $\bar{\lambda}_{h_1 H_2^\pm H_2^\pm}$ and $\bar{\lambda}_{h_1 H^{\pm\pm} H^{\pm\pm}}$. All generated points passed upon the full set of constraints at $2\sigma$. The upper panel displays the charged Higgs masses  $m_{H^\pm_1}$ (left), $m_{H^\pm_2}$ (middle) and $m_{H^{\pm\pm}}$ (right) in the $(\lambda_7,\,\,\lambda_9)$ plane. We can clearly see that $\lambda_7$ and $\lambda_9$ drastically  reduce to $0<\lambda_7<2\pi$ and $-5.25<\lambda_9<1.01$ due to the combination of the BFB conditions with the Higgs exclusion limits implemented in \texttt{HiggsTools}.  One can also notice that  $\lambda_{7}$ and $\lambda_{9}$ have opposite effects on the trilinear couplings, as illustrated by the middle and lower panels in Fig. \ref{corre_betw_coupla}: $\lambda_{7}$ increases $\bar{\lambda}_{h_1H^{\pm\pm}H^{\mp\mp}}$,$\bar{\lambda}_{h_1 H_1^\pm H_1^\pm}$,$\bar{\lambda}_{h_1 H_2^\pm H_2^\pm}$, while $\lambda_{9}$ weaken these trilinear couplings. Similarly compatibility with all the constraints oblige the parameters $\mu_{2,3}$ to undergo drastic reductions being confined within the mitigated intervals, $0\GeV< \mu_2 <8 \GeV$ and $70\GeV< \mu_3 <108 \GeV$ respectively.
		
		\begin{figure}[H]	
			\begin{minipage}{0.32\textwidth}
				\centering	
				\includegraphics[height =5cm,width=6cm]{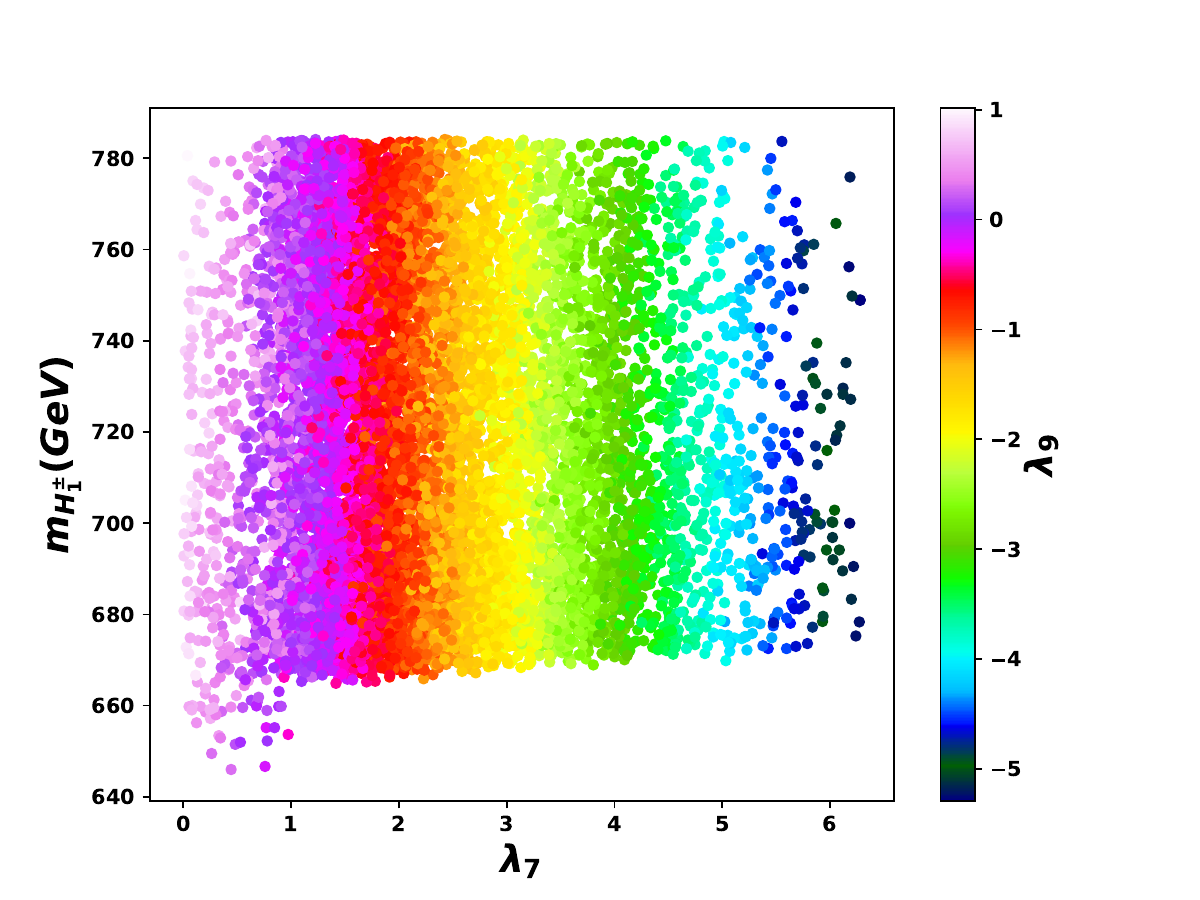}
			\end{minipage}
			\begin{minipage}{0.32\textwidth}
				\centering	
				\includegraphics[height =5cm,width=6cm]{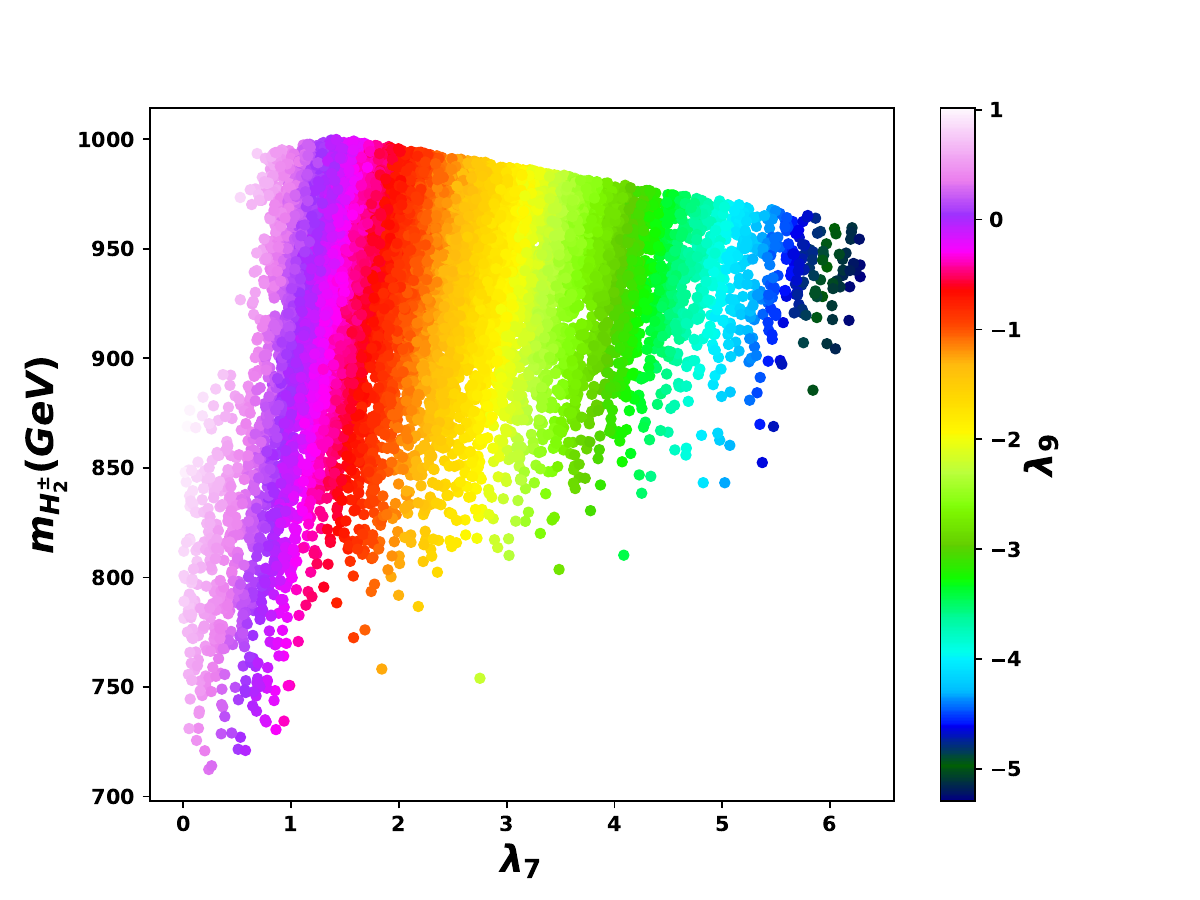}
			\end{minipage}
			\begin{minipage}{0.32\textwidth}
				\centering	
				\includegraphics[height =5cm,width=6cm]{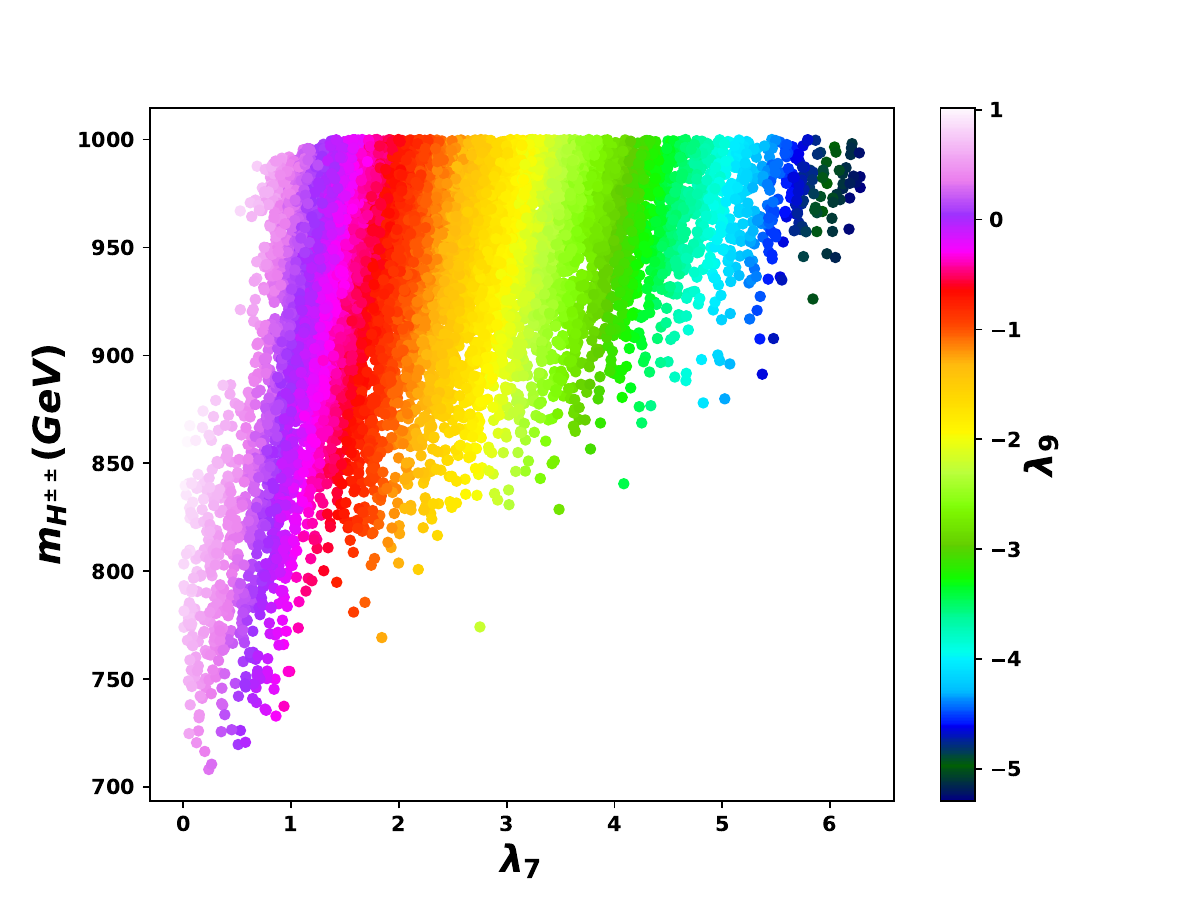} 
			\end{minipage}
			\begin{minipage}{0.32\textwidth}
				\centering	
				\includegraphics[height =5cm,width=6cm]{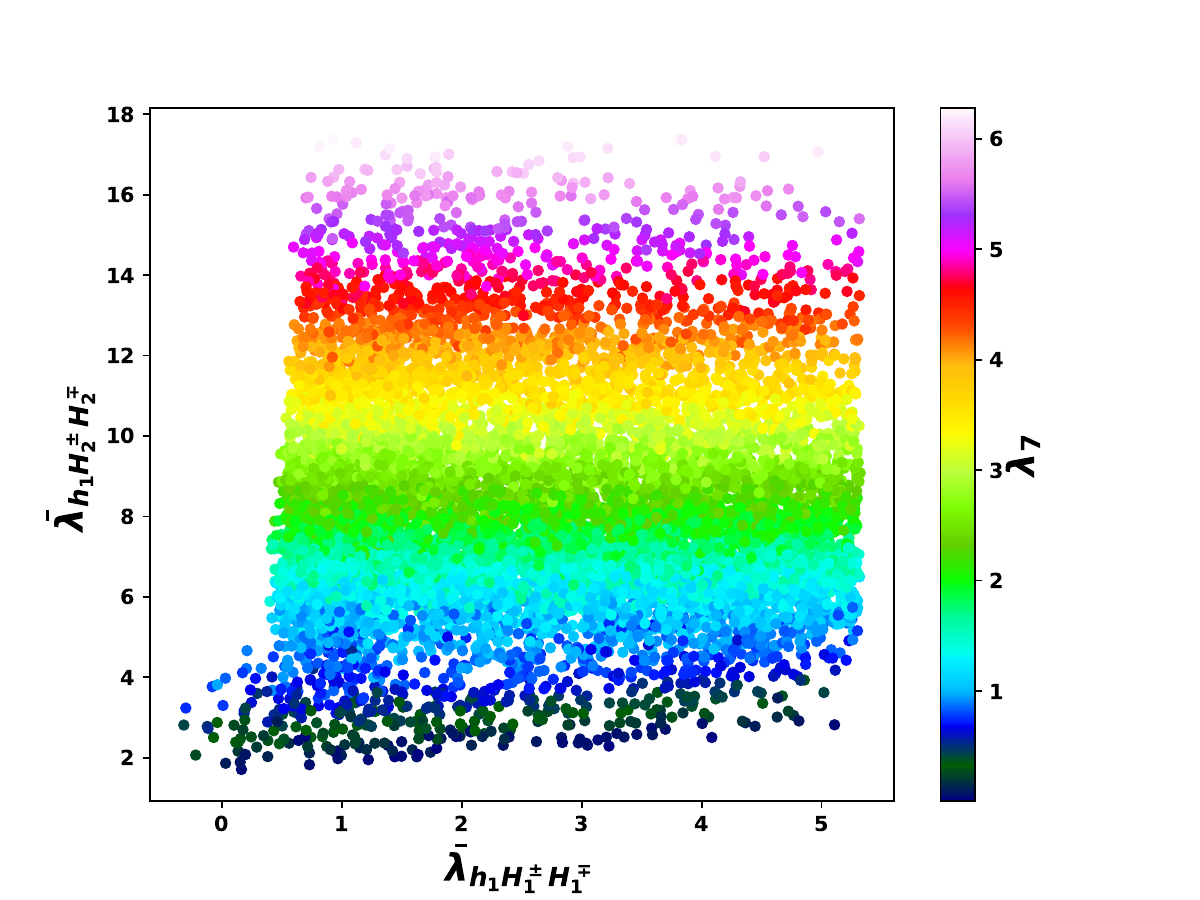}
			\end{minipage}
			\begin{minipage}{0.32\textwidth}
				\centering	
				\includegraphics[height =5cm,width=6cm]{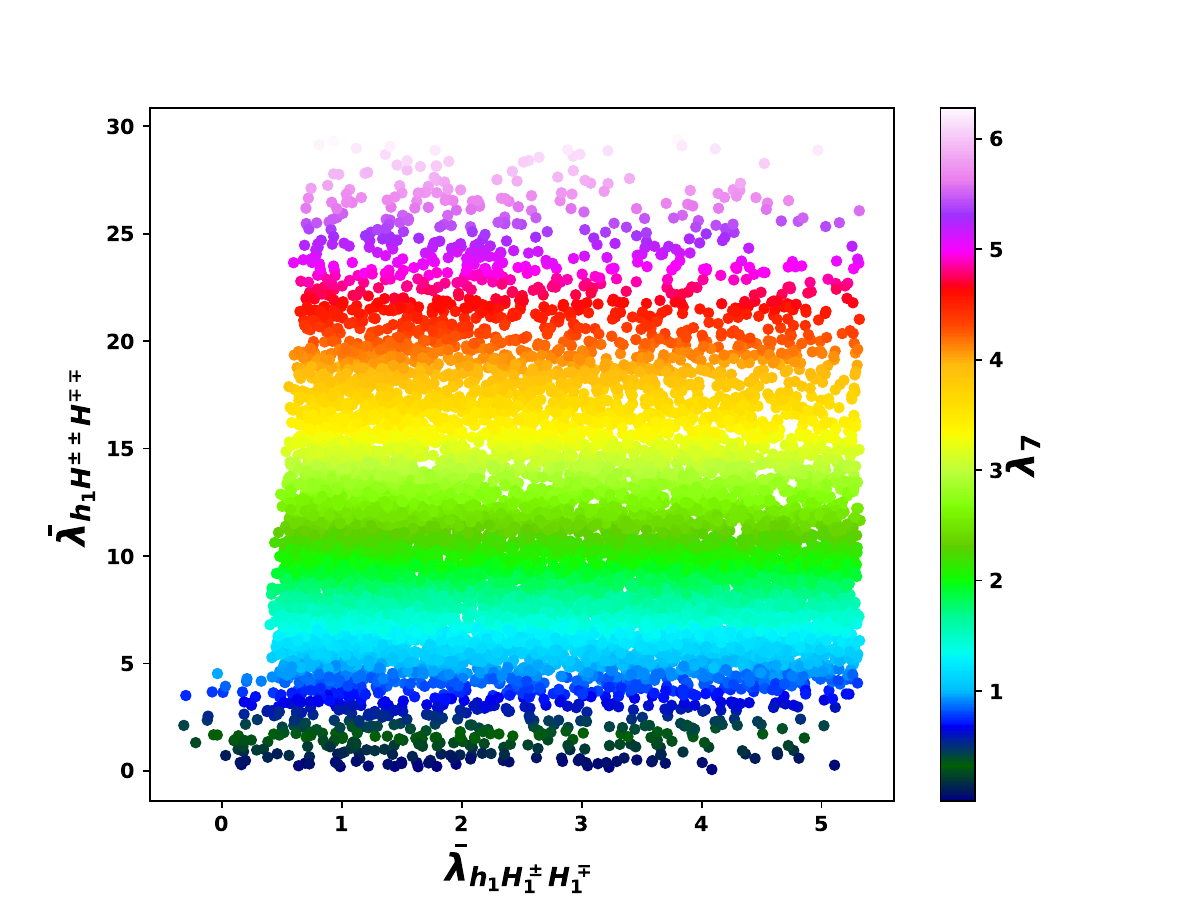}
			\end{minipage}
			\begin{minipage}{0.32\textwidth}
				\centering	
				\includegraphics[height =5cm,width=6cm]{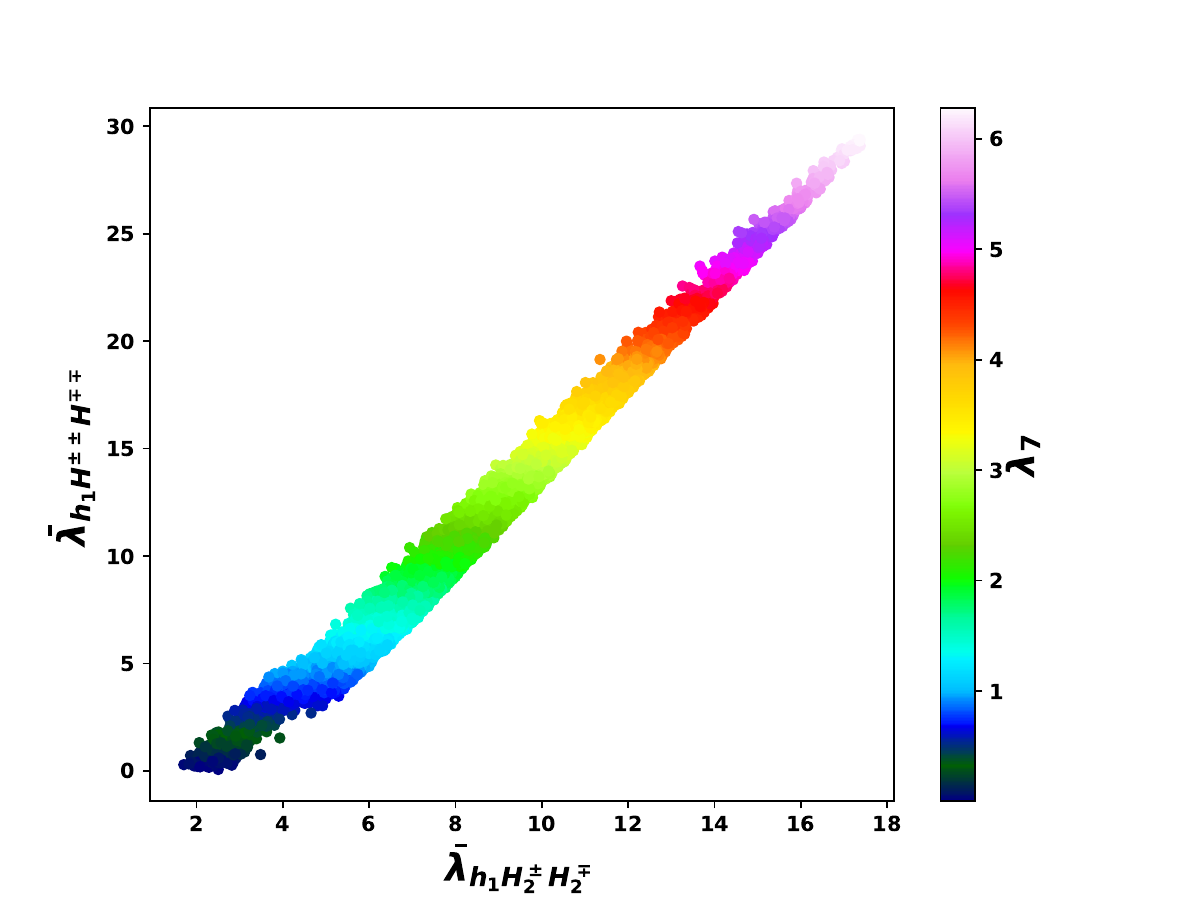} 
			\end{minipage}
			\begin{minipage}{0.32\textwidth}
				\centering	
				\includegraphics[height =5cm,width=6cm]{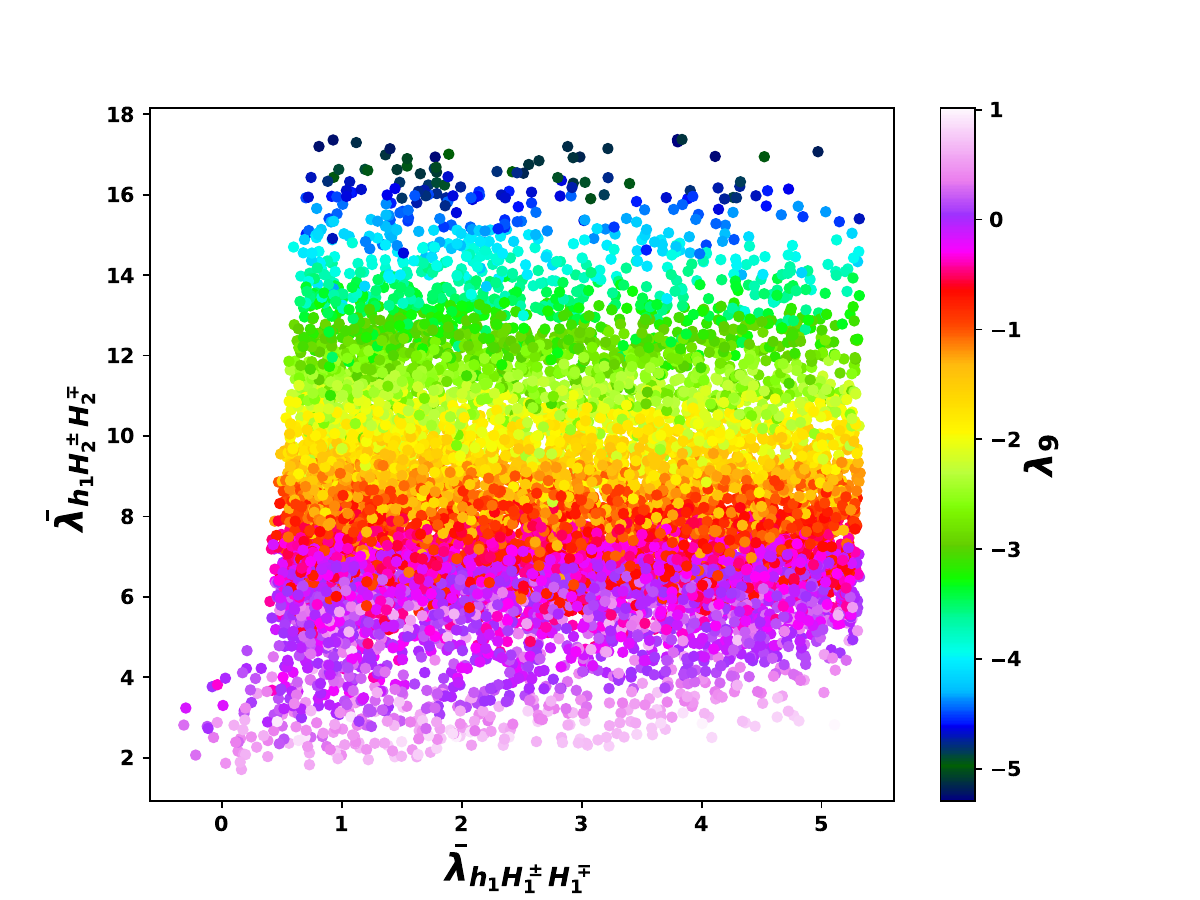}
			\end{minipage}
			\begin{minipage}{0.32\textwidth}
				\centering	
				\includegraphics[height =5cm,width=6cm]{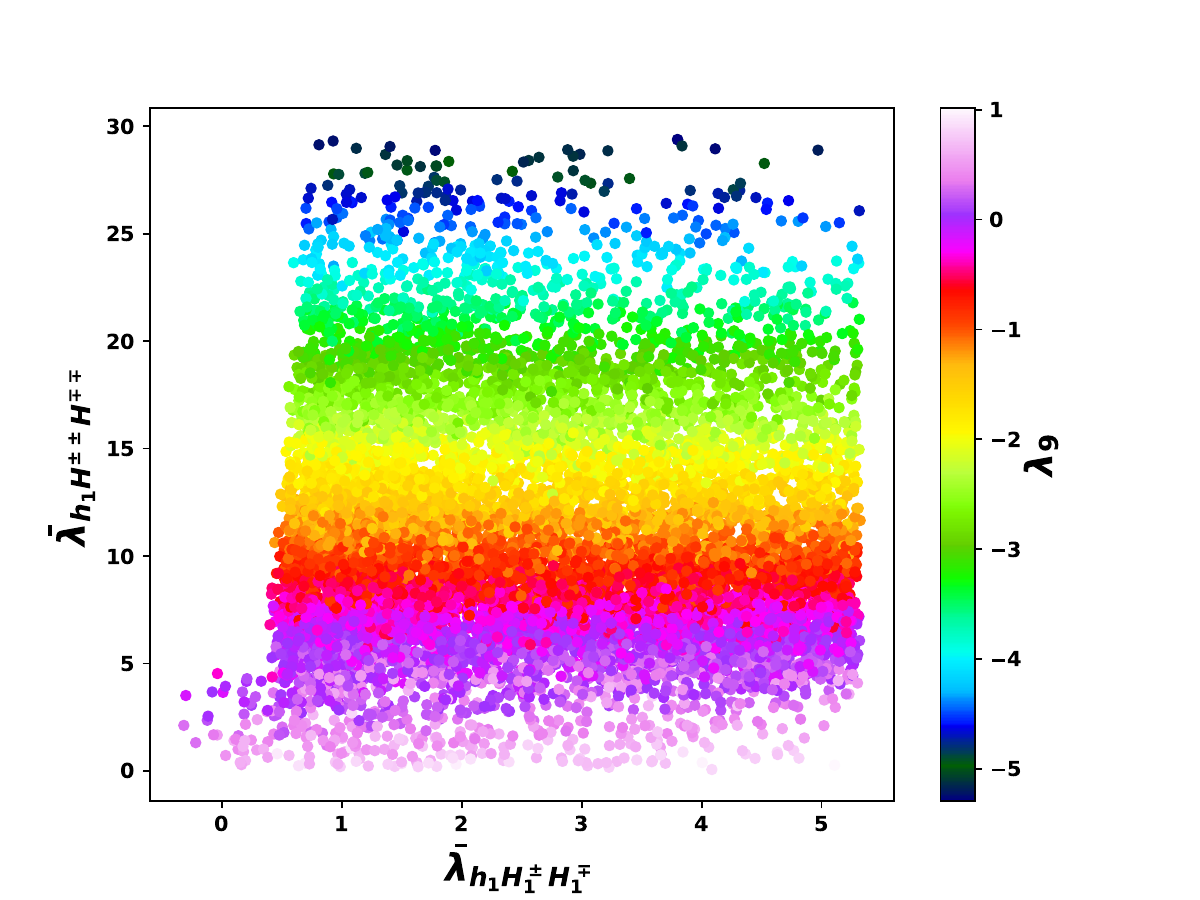}
			\end{minipage}
			\begin{minipage}{0.32\textwidth}
				\centering	
				\includegraphics[height =5cm,width=6cm]{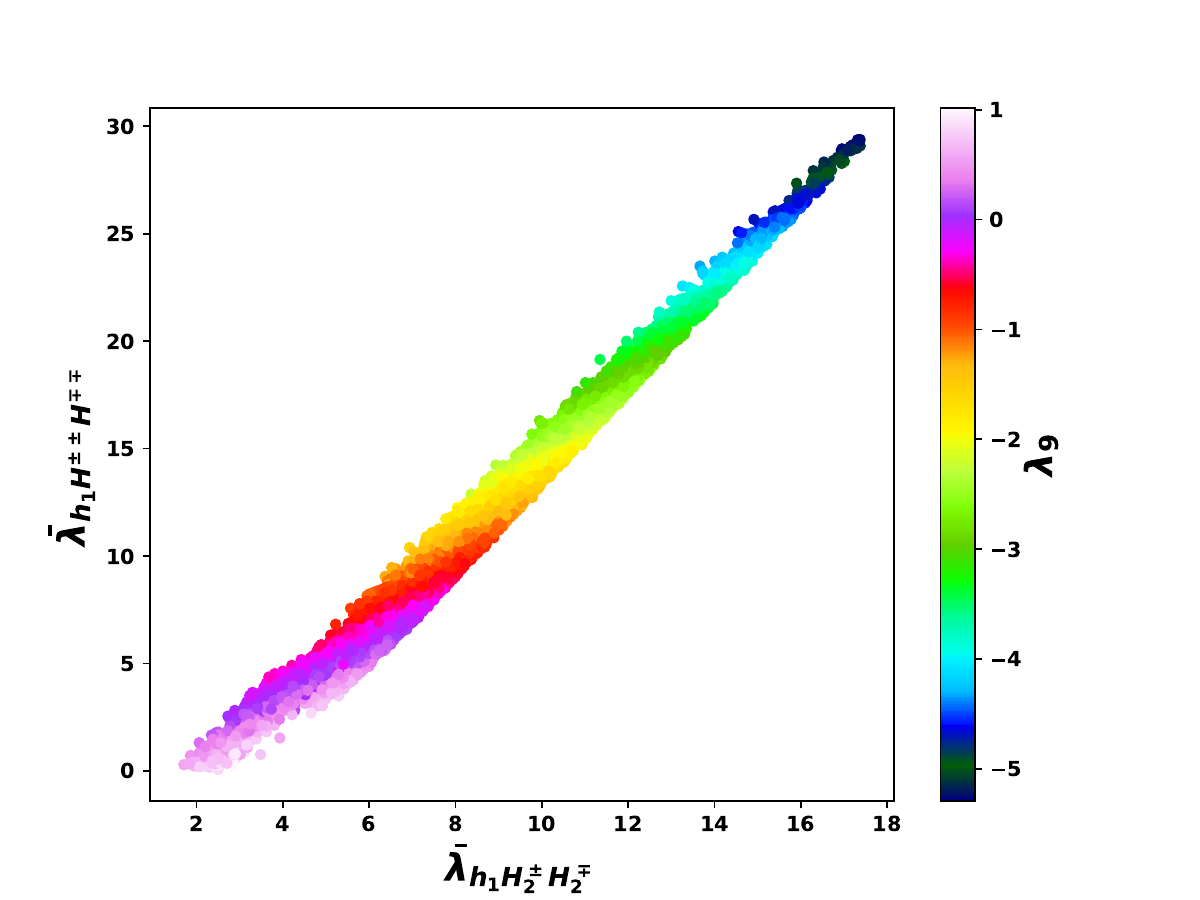} 
			\end{minipage}
			\caption[titre court]{Upper panels: The charged Higgs masses as a function of $\lambda_7$ and $\lambda_9$ with $m_{H^\pm_1}$ (left), $m_{H^\pm_2}$ (middle) and $m_{H^{\pm\pm}}$ (right). Middle panels: correlation between $\bar{\lambda}_{h_1H^\pm_1H^\mp_1}$ and $\bar{\lambda}_{h_1H^\pm_2H^\mp_2}$ (left), $\bar{\lambda}_{h_1H^\pm_1H^\mp_1}$ and $\bar{\lambda}_{h_1H^{\pm\pm}H^{\mp\mp}}$ (middle), $\bar{\lambda}_{h_1H^\pm_2H^\mp_2}$ and $\bar{\lambda}_{h_1H^{\pm\pm}H^{\mp\mp}}$ (right) as a function of $\lambda_7$. Lower panels: correlation between $\bar{\lambda}_{h_1H^\pm_1H^\mp_1}$ and $\bar{\lambda}_{h_1H^\pm_2H^\mp_2}$ (left), $\bar{\lambda}_{h_1H^\pm_1H^\mp_1}$ and $\bar{\lambda}_{h_1H^{\pm\pm}H^{\mp\mp}}$ (middle), $\bar{\lambda}_{h_1H^\pm_2H^\mp_2}$ and $\bar{\lambda}_{h_1H^{\pm\pm}H^{\mp\mp}}$ (right) as a function of $\lambda_9$. Here we assume that $\alpha_1=1.42$ and $m_{h_2}=673.30$\,\,\text{GeV}. The other inputs are the same as in (\ref{11}). The analysis is performed with  $\chi^2_{ST}$ test where only points that are within $95\%$ C.L. are considered.}
			\label{corre_betw_coupla}
		\end{figure} 
		
		\begin{figure}[H]
			\begin{minipage}{0.52\textwidth}
				\centering
				\includegraphics[height =6cm,width=7cm]{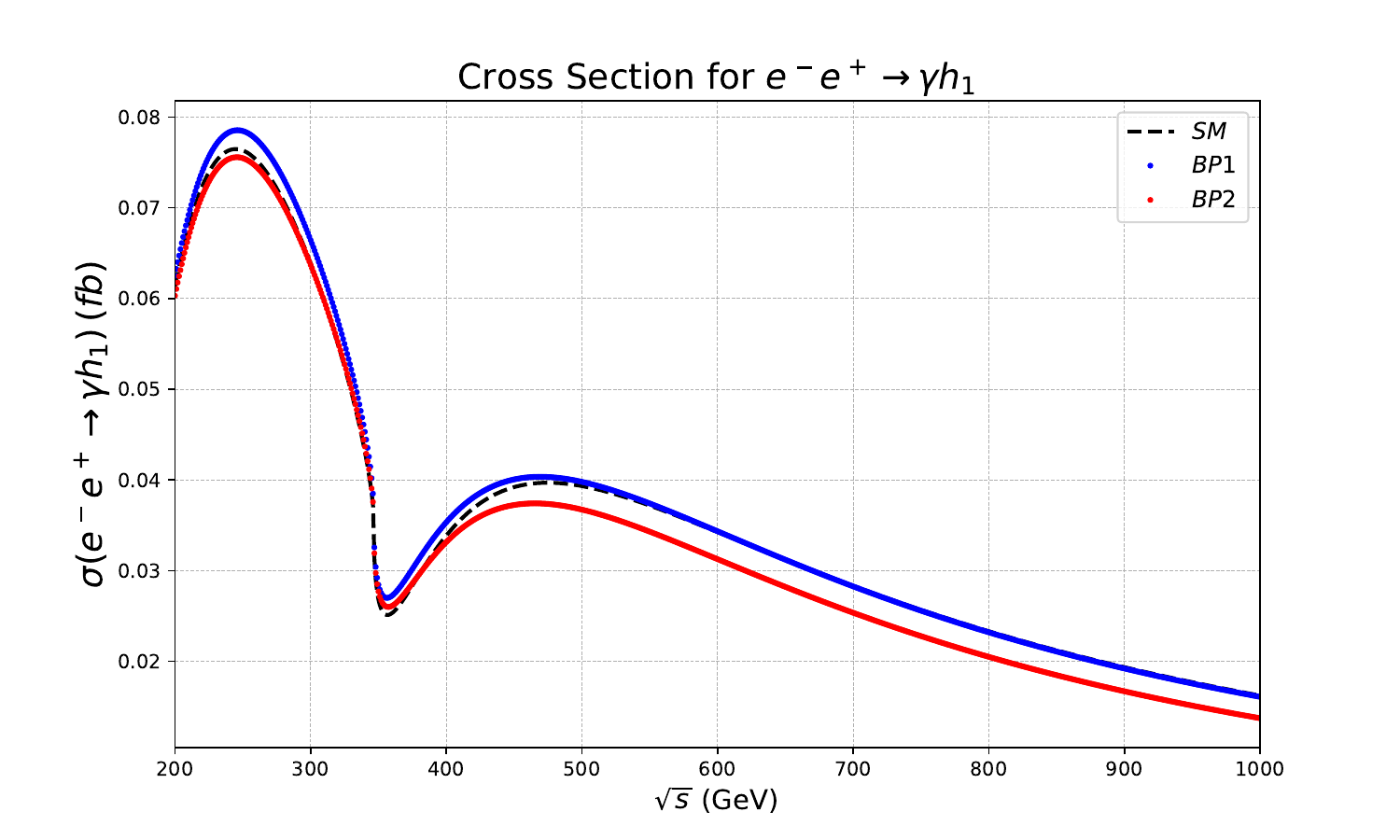}
			\end{minipage}
			\begin{minipage}{0.52\textwidth}
				\centering
				\includegraphics[height =6cm,width=7cm]{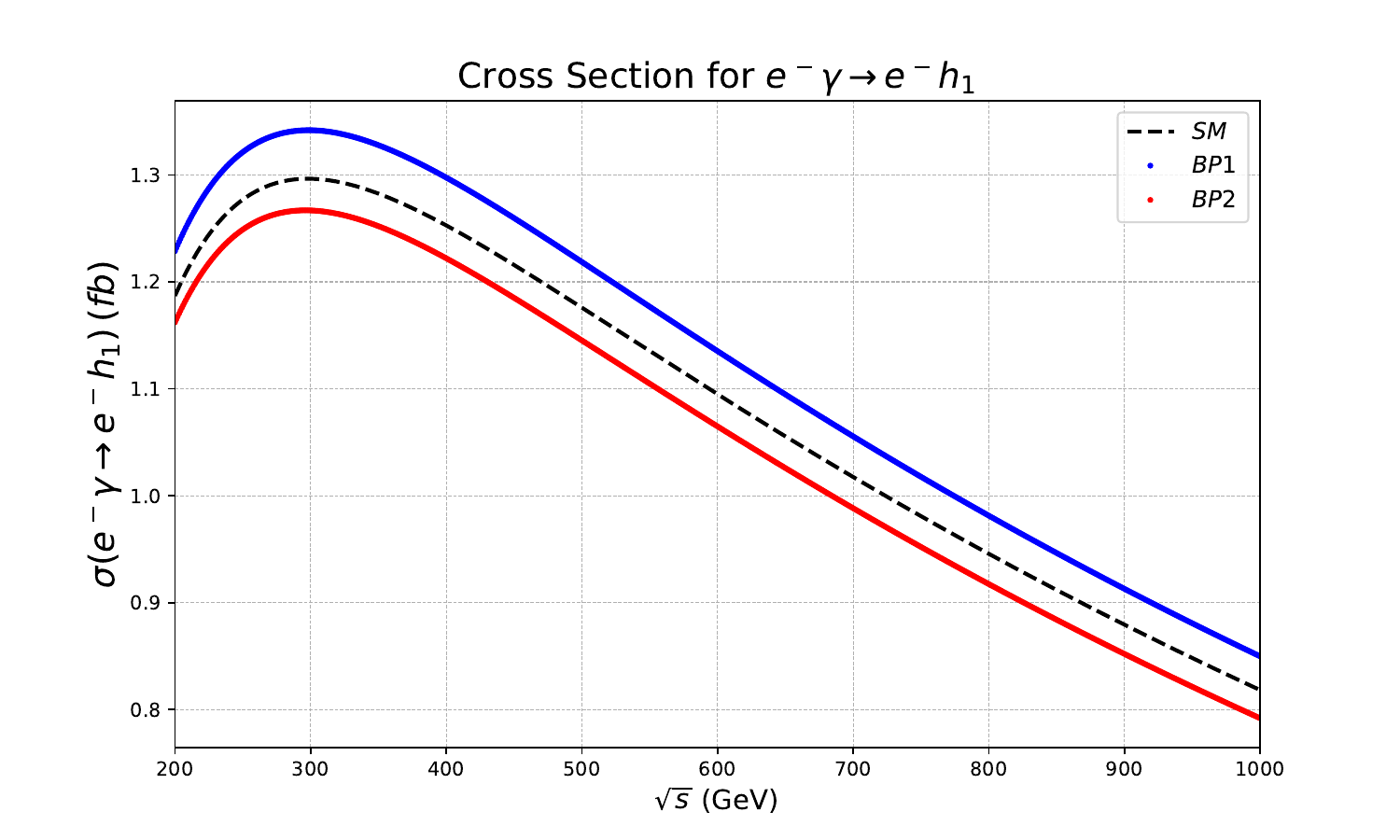}
			\end{minipage}
			\caption[titre court]{Total unpolarized cross-sections in fb for $e^+e^-\to \gamma h_1$ (left) and $e^- \gamma \to e^- h_1$ (right) as a function of center-of-mass energy, in two benchmark scenarios. Here we assume that $\alpha_1=1.42$ and $m_{h_2}=673.30$\,\,\text{GeV}. The other inputs are dispayed in Table \ref{tab:bp_nonalignment}.}
			\label{fig:6}
		\end{figure} 
		
		\paragraph*{}
	Fig. \ref{fig:6} dispalys the unpolarized cross sections of $e^- e^+ \to h_1 \gamma$ and $e^- \gamma \to e^- h_1$ in 2HDMcT as a function of $\sqrt{s}$ in the range $200$ to $1000$ GeV.  These cross sections at the linear collider are analyzed in two benchmark scenarios illustrated in Table \ref{tab:bp_nonalignment}. We see that the $\sigma (e^- e^+ \to h_1 \gamma)$  is significantly enhanced near $\sqrt{s}= 250$ GeV, however as $\sqrt{s}$ increases the cross section decreases until reaching  a local minimum around $t\bar{t}$ threshold $\sqrt{s}\approx346$ GeV. This is a general trend of the cross section seen in SM, and in many extended Higgs sector models  \cite{Arhrib:2014pva,Rahili:2019ixf}.  In comparison to SM, the interference between the SM and the new $2HDMcT$ contributions can either be constructive  or destructive, thus implying  an increase or a decrease of the cross section. From Eq. $6$ in our paper, on can readily see  that the cross section for $\sigma (e^- e^+ \to h_1 \gamma)$ scales like $1/s$ for large $\sqrt{s}$, hence explaining the decrease in this cross section at high $\sqrt{s}$.  \\ 
	As for  $\sigma (e^- \gamma \to e^- h_1)$ decay, it shows a slower decrease at high values of energy, due essentially to the t-channel contribution. At this point, we must stress that the  interference between $2HDMcT$ and  SM contributions  is sensitive to  the couplings $\bar{\lambda}_{h_1H^{\pm\pm}H^{\mp\mp}}$, $\bar{\lambda}_{h_1 H_1^\pm H_1^\pm}$, and $\bar{\lambda}_{h_1 H_2^\pm H_2^\pm}$. From Eqs $(14,15,16)$. we see that thess trilinear Higgs couplings  are experssed in terms of several potential parameters. However our analysis finds that $e^- e^+ \to h_1 \gamma$ and $e^- \gamma \to e^- h_1$ cross sections are mainly dependent on $\lambda_{3}$, $\lambda_{7}$ and $\lambda_{9}$, with a noticeable sensitivity $\alpha_1$. 
		\begin{table}[!h]
			\caption{Benchmark scenarios used in Fig. \ref{fig:6} where the generated points pass upon the full set of constraints.}
			\centering
			\begin{tabular}{c|c|c|c|c|c|c|c|c|c|c}
				\hline
				Bench. & $\lambda_1$ & $\lambda_3$ & $\lambda_4$ & $\lambda_6$ &  $\lambda_7$ & $\lambda_8$ & $\lambda_9$ & $\bar{\lambda}_8$ &  $\bar{\lambda}_9$ &      \\ \hline 
				\hline
				BP1 & $1.31$ & $-0.12$ & $1.70$ & $0.64$ & $0.01$ & $0.28$ & $0.38$ & $2.33$ & $0.54$ &  \\ 
				BP2 & $1.31$ & $0.80$ & $3.06$ & $6.04$ & $2.15$ & $-2.72$ & $0.47$ & $-1.10$ & $3.13$ &  \\
				\hline 
				%1 & 2 & 3 \\
				\hline
			\end{tabular}
			\label{tab:bp_nonalignment}
		\end{table}
	\begin{figure}[H]
		\centering	
		\begin{minipage}{0.32\textwidth}
			\includegraphics[height =5cm,width=5.6cm]{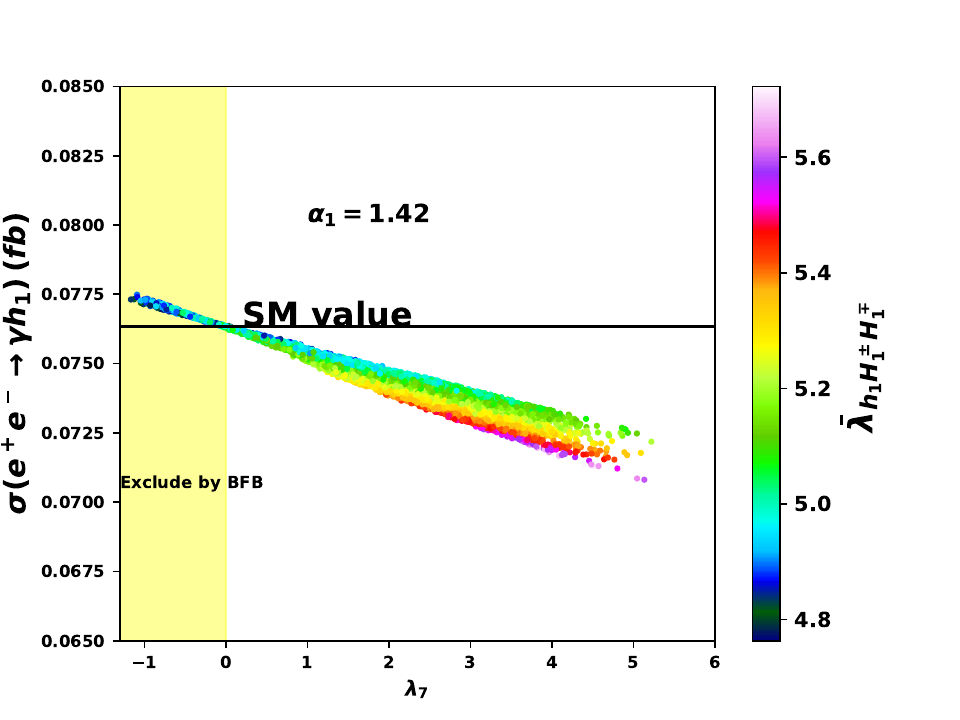}
		\end{minipage}
		\begin{minipage}{0.32\textwidth}
			\includegraphics[height =5cm,width=5.6cm]{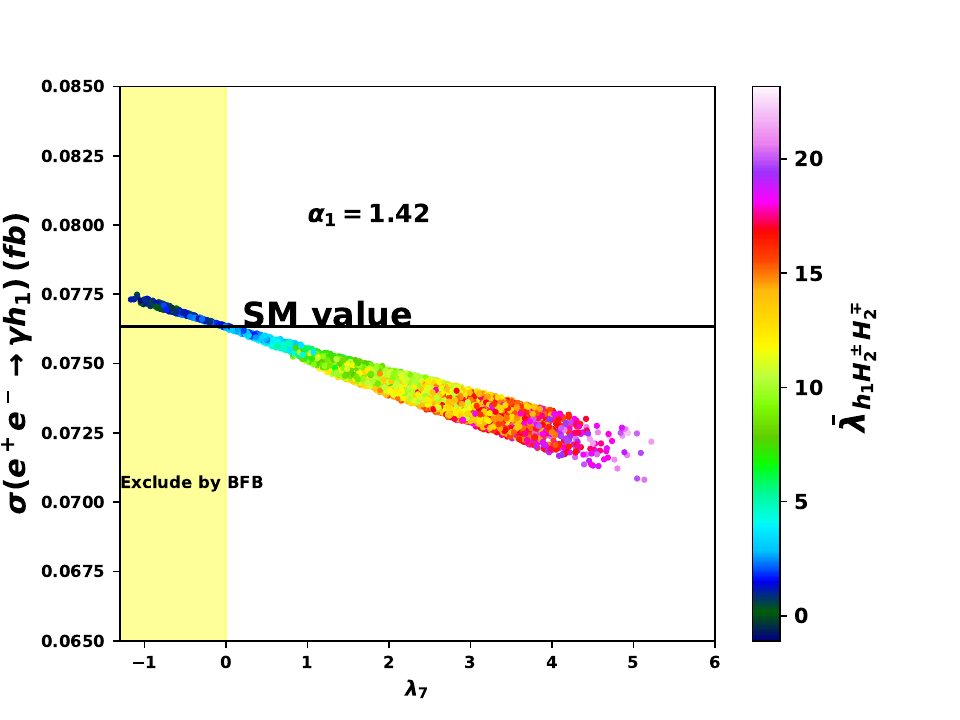}
		\end{minipage}
		\begin{minipage}{0.32\textwidth}
			\includegraphics[height =5cm,width=5.6cm]{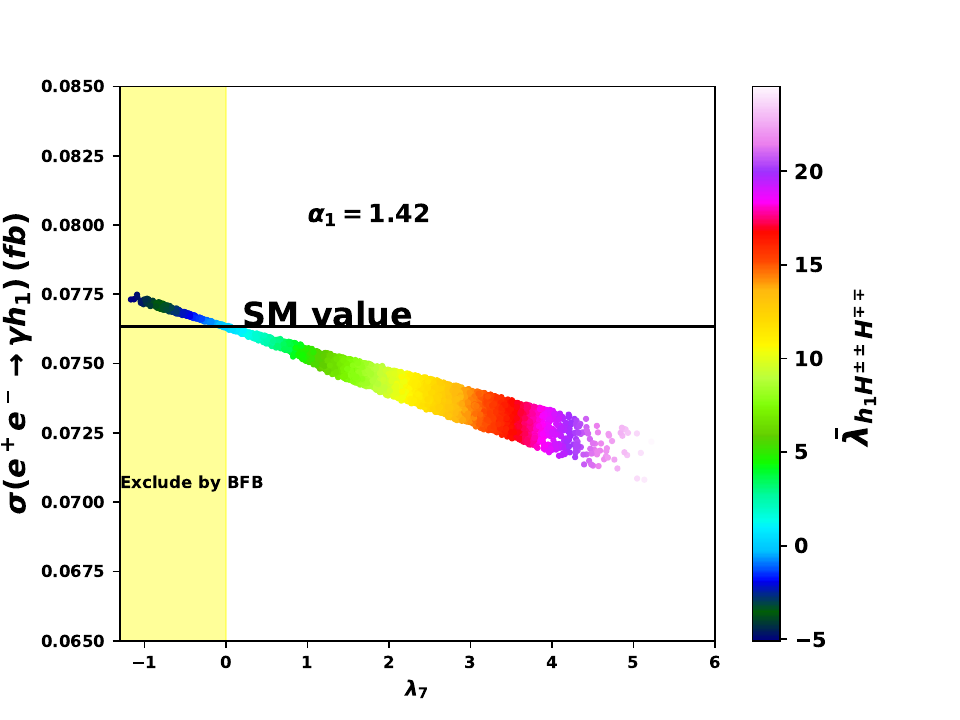} 
		\end{minipage}
		\begin{minipage}{0.32\textwidth}
			\includegraphics[height =5cm,width=5.6cm]{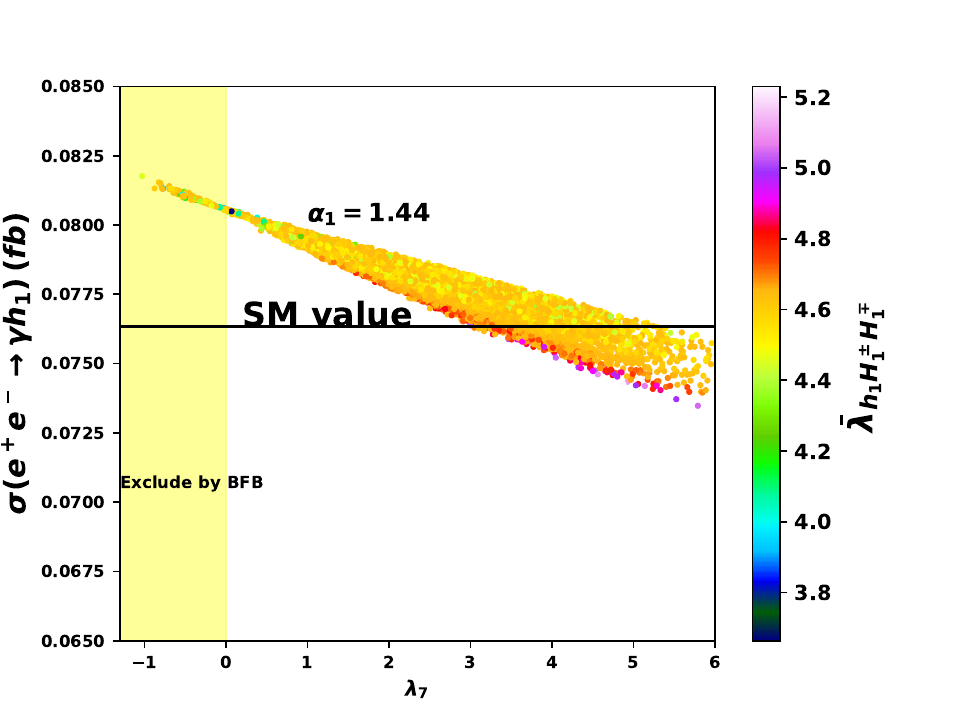}
		\end{minipage}
		\begin{minipage}{0.32\textwidth}
			\includegraphics[height =5cm,width=5.6cm]{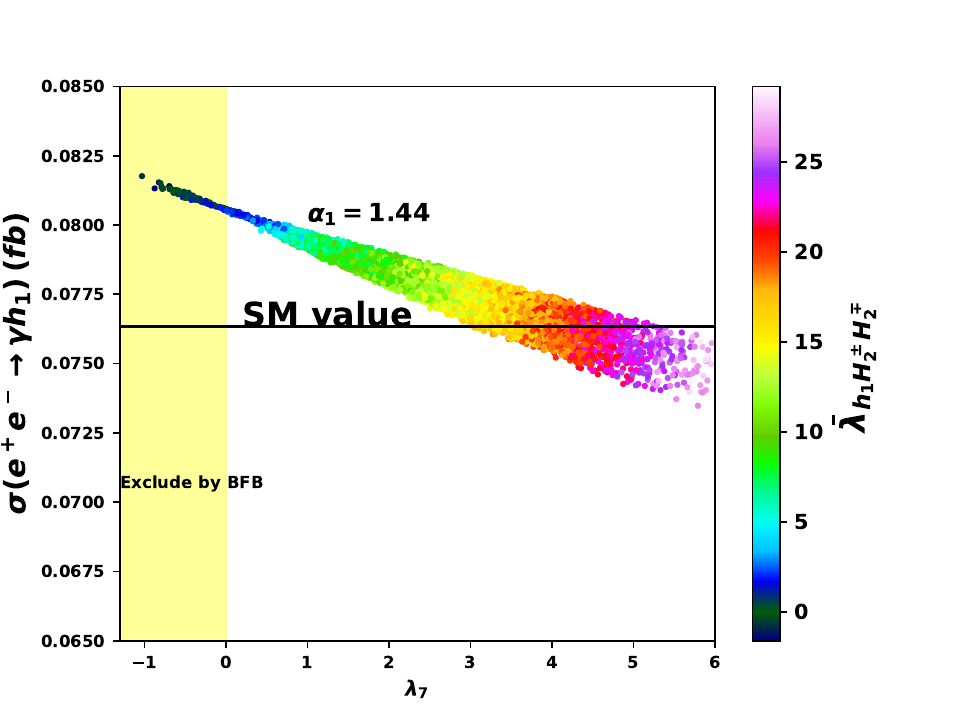}
		\end{minipage}
		\begin{minipage}{0.32\textwidth}
			\includegraphics[height =5cm,width=5.6cm]{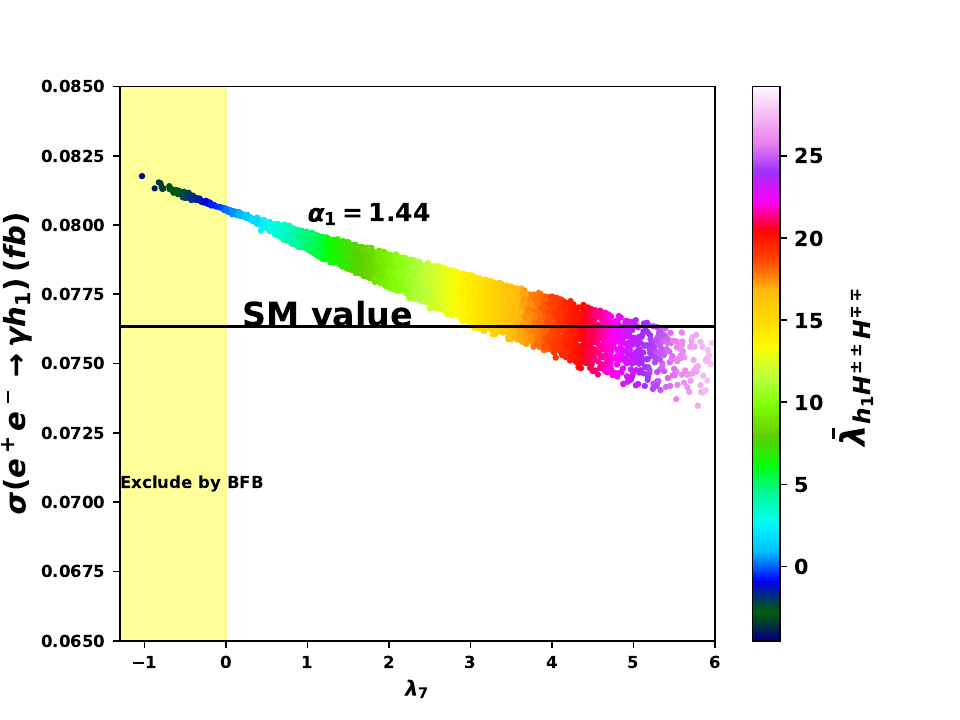} 
		\end{minipage}
		\caption[titre court]{The cross-section  $\sigma(e^+e^- \to \gamma h_1)$ is depicted as a function of the parameter $\lambda_7 $ for two values of the parameter $\alpha_1$: $\alpha_1=1.42$ (upper panels), and $\alpha_1=1.44$ (lower panels). The plotted points pass upon all constraints.
		We used : $\lambda_1= 1.31$, $\lambda_3= 6.15$, $\lambda_4= -3.85$ and $m_{h_2}=673.30$ GeV.  The other inputs are the same as in (\ref{11}).  The color coding illustrates  the variation of the trilinear couplings $\bar{\lambda}_{h_1H^\pm_1H^\mp_1}$(left), $\bar{\lambda}_{h_1H^\pm_2H^\mp_2}$(middle)  and 
		$\bar{\lambda}_{h_1 H^{\pm\pm}H^{\mp\mp}}$ (right panel)}.
		\label{fig5}
	\end{figure} 
	
		Next, we first aim to see whether the diphoton signal strength $R_{\gamma\gamma}(h_1)$ agrees  with the experimental data within  the model's parameter space, knowing that, unlike $h_1 \to \gamma Z$ process, the decay channel $h_1 \to \gamma\gamma$, has been measured with high precision \cite{ATLAS:2022tnm, ATLAS:2020qcv, CMS:2022ahq, CMS:2023mku, ATLAS:2023wqy,ATLAS:2022vkf,CMS:2022dwd}. Also,  we want to analyze to what extent the cross-section $\sigma(e^+e^- \to \gamma h_1)$ behavior is  affected either by the $2HDMcT$ charged Higgs bosons or by the potential parameters. Fig. \ref{fig5} and Fig. \ref{fig:07} illustrate the variation of these observables as a function of the parameter $\lambda_7$,  and the trilinear couplings $\bar{\lambda}_{h_1 H^{+}_1H^{-}_1}$, $\bar{\lambda}_{h_1 H^{+}_2H^{-}_2}$ and $\bar{\lambda}_{h_1 H^{++}H^{--}}$. It is generally observed that an increase in $\lambda_7$ results in a detrimental effect on both $R_{\gamma\gamma}$ and $\sigma(e^+e^- \to \gamma h_1)$. Similar trend is seen if these observables are plotted versus  $\lambda_{3}$.
		
       In Fig. \ref{fig5} we plot the cross section of $e^+e^- \to \gamma h_1$  as a function of the  parameter $\lambda_7$, and the trilinear Higgs couplings for two nearly degenerate values of the parameter $\alpha_1$. From the upper panels, with $\alpha_1=1.42$, we clearly see that $\sigma(e^+e^- \to \gamma h_1)$ undergoes a notable increase and even exceeds the SM prediction if $\lambda_7<0$, a region  excluded by the $BFB$ constraints. In this scenario, because of $BFB$ conditions, the contributions of  the parameters $\lambda_{6,7,8,9, \bar{8},\bar{9}}$ are rather predominantly detrimental,  inducing a decrease in $\sigma(e^+e^-\to\gamma h_1)$. More specifically this depreciation stems  particularly from the relations, $\lambda_7>0$ and $\lambda_7+\lambda_9 > 0$, with  $\lambda_7$ and $\lambda_9$ exhibiting contrasting effects. However,  if we consider a slightly larger value of the parameter $\alpha_1=1.44$, as illustrated by the lower panels, the cross section experience a significant enhancement and can well surpass that of the SM up to $8.1 \times 10^{-2}$  fb, while still satisfying BFB conditions. This underscores the strong sensitivity to $\alpha_1$. In addition, as a byproduct,  our analysis also show that the charged Higgs masses $m_{H^\pm_1}$, $m_{H^\pm_2}$ and $m_{H^{\pm\pm}}$ are compelled to lie within the very stringent intervals $[785, 790]$ GeV, $[807, 990]$ GeV and $[810 , 995]$ GeV respectively, confirming the predictions reported in ~\cite{Ouazghour:2023eqr}.\\
       
        Next we analyze the signal strengths $R_{\gamma\gamma}$, $R_{\gamma Z}$ and the ratio $R_{\gamma h_1}$ within the model parameter space. First,  we plot in Fig.\ref{fig:07} the signal strength of the Higgs  to  diphoton decay as a function of the parameter $\lambda_7$, so we observe that $R_{\gamma\gamma}$ is clearly consistent with the measured signal strength \cite{ATLAS:2022tnm, ATLAS:2020qcv, CMS:2022ahq, CMS:2023mku, ATLAS:2023wqy,ATLAS:2022vkf,CMS:2022dwd} at $1\sigma$.
		\begin{figure}[H]
			\begin{minipage}{0.32\textwidth}
				\centering
				\includegraphics[height =5.5cm,width=5.5cm]{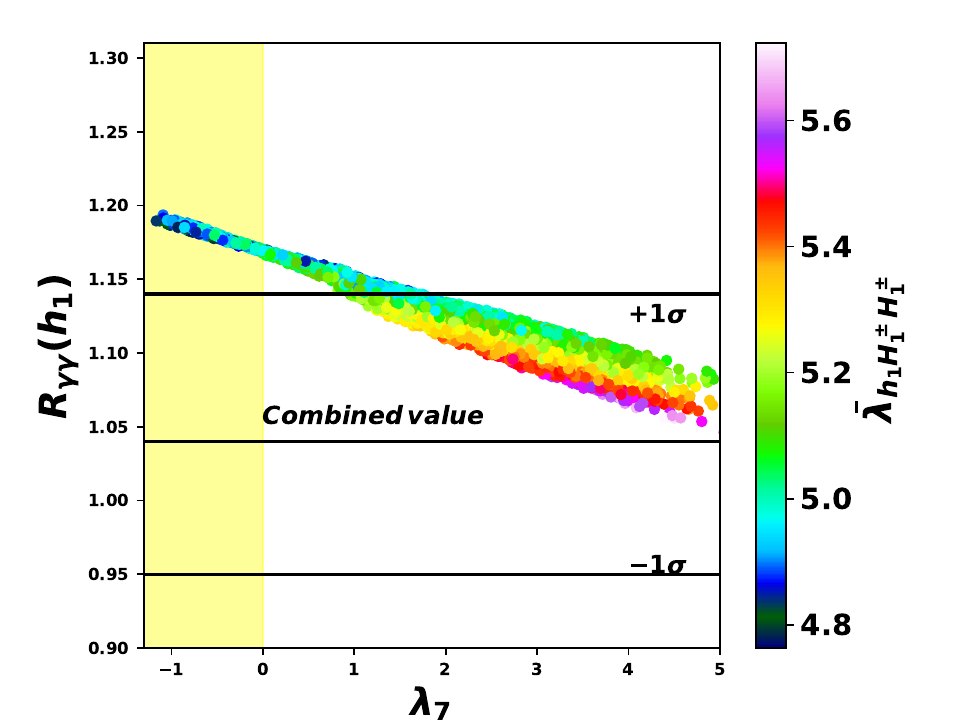}
			\end{minipage}
			\begin{minipage}{0.32\textwidth}
				\centering
				\includegraphics[height =5.5cm,width=5.5cm]{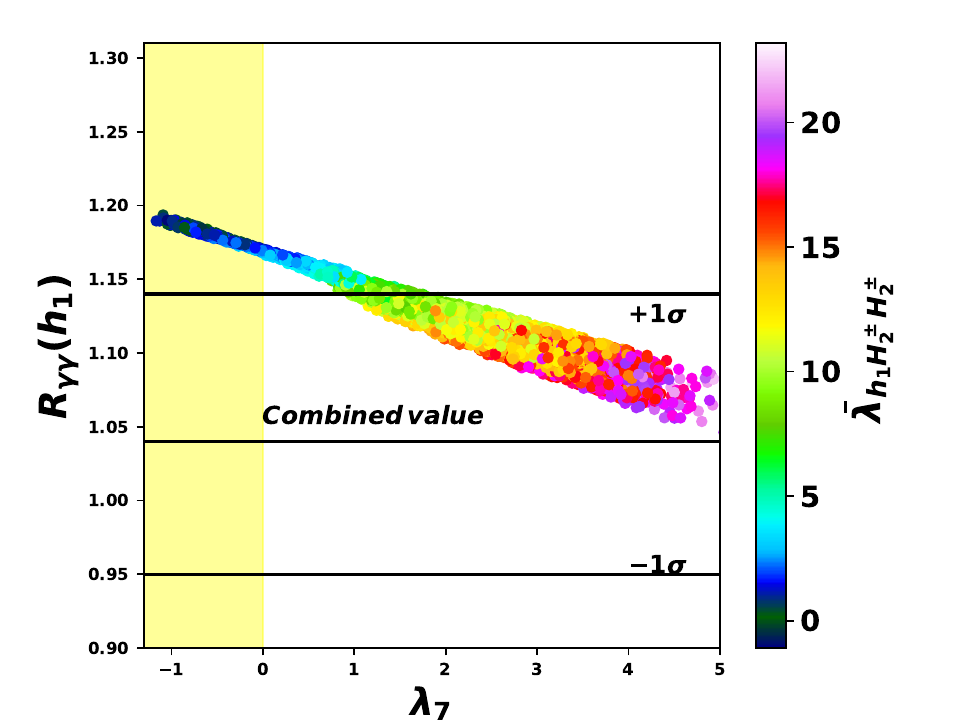}
			\end{minipage}
			\begin{minipage}{0.32\textwidth}
				\centering
				\includegraphics[height =5.5cm,width=5.5cm]{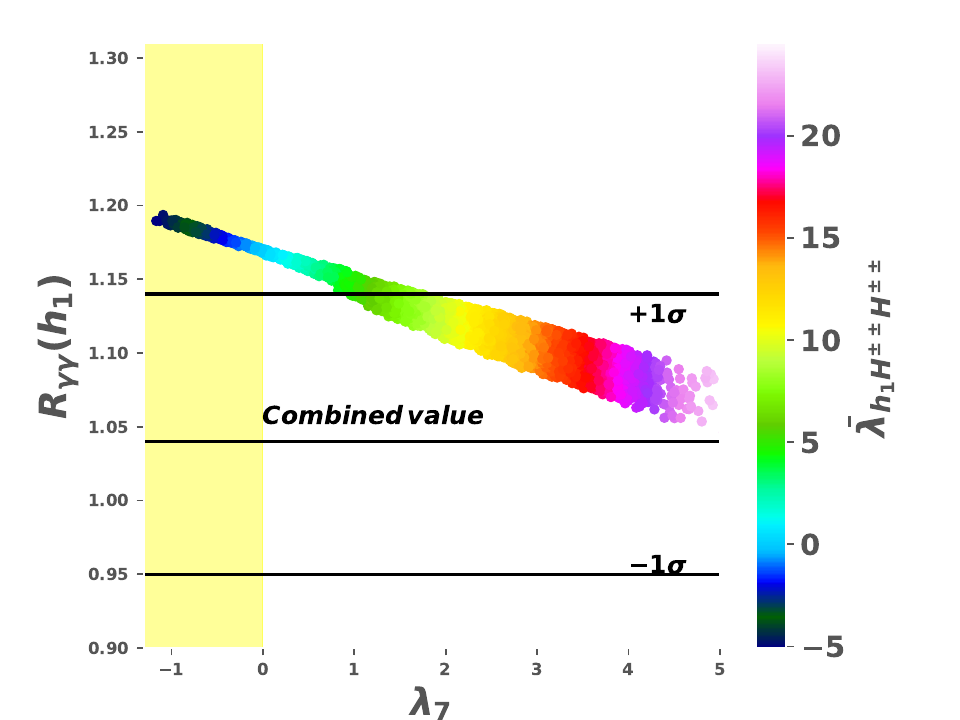}
			\end{minipage}
		
			\caption[titre court]{The signal strength $R_{\gamma\gamma}(h_1)$ as a function of  $\lambda_7 $. The plotted points passed upon all constraints. The color coding shows the variation of the trilinear coupling $\bar{\lambda}_{h_1H^\pm_1H^\mp_1}$(left), $\bar{\lambda}_{h_1H^\pm_2H^\mp_2}$(middle) and 
		$\bar{\lambda}_{h_1 H^{\pm\pm}H^{\mp\mp}}$ (right panel). The displayed horizontal lines denote the central and $\pm1\sigma$ diphoton signal strength values reported by ATLAS at $13$ TeV \cite{ATLAS:2022tnm,ATLAS:2022vkf}. We used the following inputs: $\alpha_1=1.42$, $\lambda_1= 1.31$, $\lambda_3= 6.15$, $\lambda_4= -3.85$ and $m_{h_2}=673.30$ GeV. The other inputs are the same as in (\ref{11}), while the error for $\chi^2$ fit is 95\% C.L.}
			\label{fig:07}
		\end{figure}
        % .\par
        Then, we investigate how $R_{\gamma h_1}$ behaves with respect to $R_{\gamma \gamma }$  and $R_{\gamma Z }$ and whether these observables are correlated. One can readily see from Fig. \ref{fig:101} a  clear correlation with these two signal strengths. Additionally, we also analyzed the incidence of the doubly charged Higgs on these loop induced processes, via its trilinear Higgs couplings $\bar{\lambda}_{h_1H^{\pm\pm} H^{\mp\mp}}$}. Interestingly the results  indicate that the lighter it is, the more significant the enhancement in $\sigma(e^+e^- \to \gamma h_1)$,  as in $R_{\gamma h_1}$ and $R_{\gamma Z}$.
		\begin{figure}[H]
			\begin{minipage}{0.52\textwidth}
				\centering	
				\includegraphics[height =6cm,width=6cm]{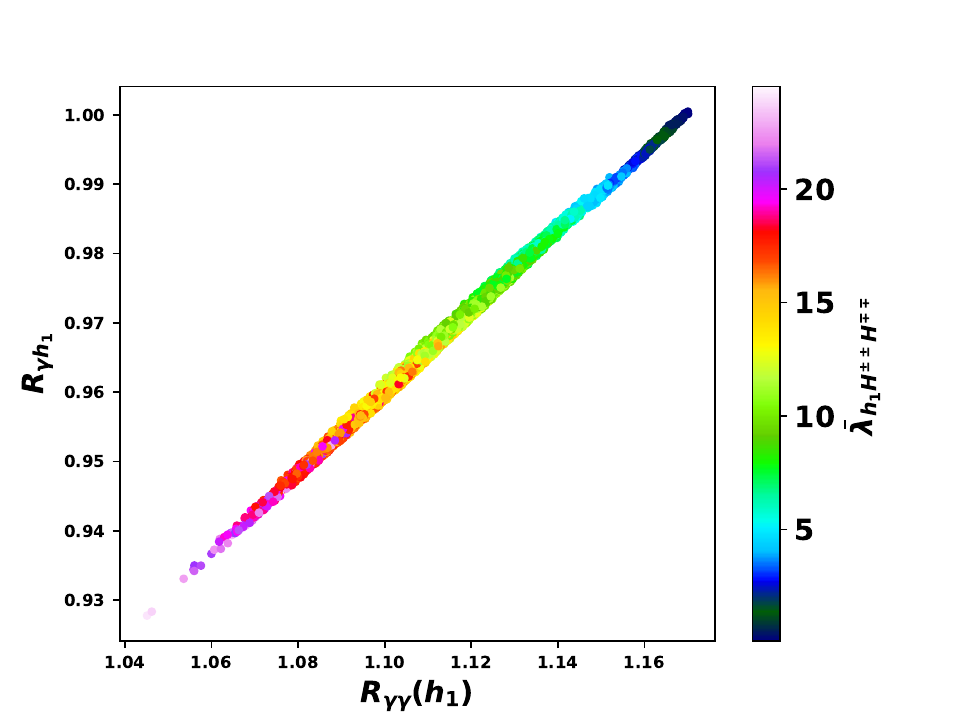}
			\end{minipage}
			\begin{minipage}{0.52\textwidth}
				\centering	
				\includegraphics[height =6cm,width=6cm]{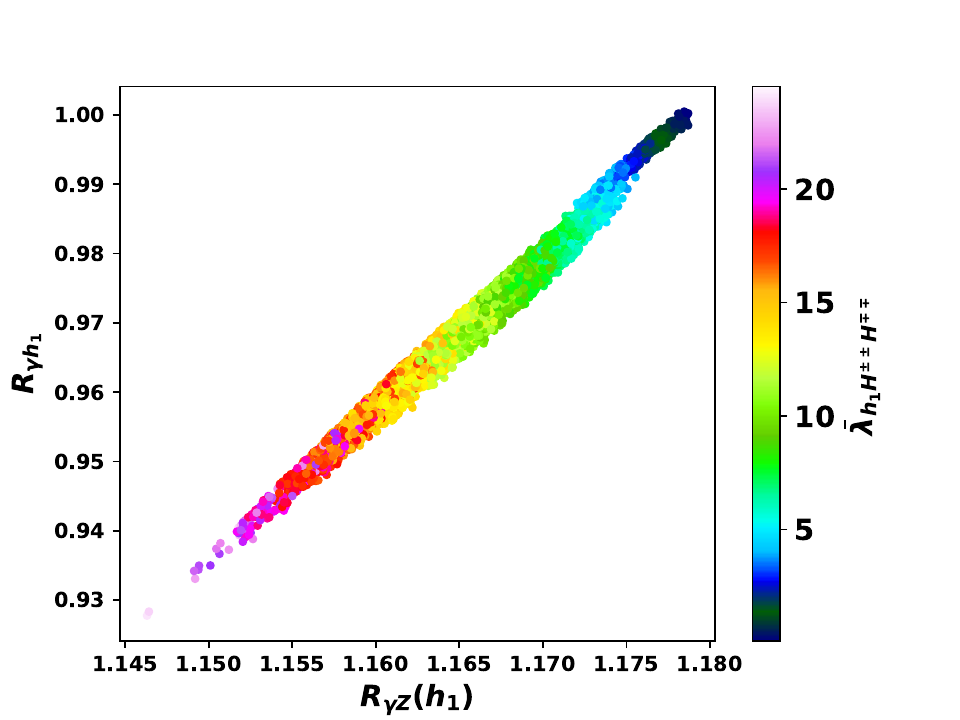}
			\end{minipage}
			\caption[titre court]{$R_{\gamma h_1}$ correlation with $R_{\gamma\gamma}(h_1)$ (left panels) and $R_{\gamma Z}(h_1)$  (right panels) in $2HDMcT$. Color coding shows the variation of $\bar{\lambda}_{h_1 H^{\pm\pm}H^{\mp\mp}}$. The other inputs are the same as in Fig. \ref{fig:07}.}
			\label{fig:101}
		\end{figure} 
			
		\section{CONCLUSION}
		\label{conlusion}
The future $e^-e^+$ colliders are anticipated to play a vital role in understanding the nature of the Higgs boson and its coupling to the $SM$ particles with unprecedented precision. 
In the present paper, we have studied the one-loop processes $e^+e^- \to \gamma h_1$ and $e^-\gamma \to e^-h_1$ with a more focus on $e^+e^- \to \gamma h_1$ in the framework of  $2HDMcT$ at the $e^-e^+$ colliders, where the $h_1$ Higgs boson is chosen to replicate the observed Higgs. Within the 2HDMcT parameter space delimited by a full set of constraints, we have shown that the 2HDMcT charged scalars can substantially alter the cross sections $\sigma(e^+e^-\to\gamma h_1)$ and $\sigma(e^-\gamma \to e^-h_1)$. Our analysis also observed that these observables are strongly dependent on the model parameters, especially $\alpha_1$, $\lambda_{7}$, $\lambda_{9}$, and the trilinear Higgs couplings, with a particularly strong sensitivity to $\alpha_1$. These parameters conspire to induce significant contributions to $\sigma (e^+e^- \to h_1\gamma)$, up to $8.1 \times 10^{-2}$ fb thus exceeding  the $SM$ prediction.

Furthermore, the charged scalars  $H_1^\pm$, $H_2^\pm$ and $H^{\pm\pm}$ also affected  the ratio $R_{h_1 \gamma }$ and the signal strength $R_{Z \gamma }$, while  $R_{\gamma \gamma}$ being consistent with the measured signal strength at $1\sigma$. At last, our anlysis showed  that these three observables are entirely correlated with each other.
\renewcommand\thesection{\Alph{section}}
\setcounter{section}{0}

	\begin{center}
		{\huge{Appendices}}
	\end{center}
	\begin{center}
	\section{Decay width functions}
		\label{Appendix_A}
	\end{center}
	Here we present the definition of the functions appearing in the Higgs decay widths.
	\begin{align}
	A_{1/2}^{h_1}(\tau) & = 2 [\tau +(\tau -1)f(\tau)]\, \tau^{-2}  \nonumber \\   
	A_1^{h_1}(\tau) & = - [2\tau^2 +3\tau+3(2\tau -1)f(\tau)]\, \tau^{-2} \nonumber \\
	A_{0}^{h_1}(\tau) & = - [\tau -f(\tau)]\, \tau^{-2}
\end{align}
\noindent
The scaling variables are defined as $\tau_i=M^2_{\Phi}/4M^2_i$, where $M_i$ represents the loop mass.

\noindent
The function $f(\tau)$ is defined as,
	\begin{eqnarray}
		f(\tau)=\left\{
		\begin{array}{ll}  \displaystyle
			\arcsin^2\sqrt{\tau} & \tau\leq 1 \\
			\displaystyle -\frac{1}{4}\left[ \log\frac{1+\sqrt{1-\tau^{-1}}}
			{1-\sqrt{1-\tau^{-1}}}-i\pi \right]^2 \hspace{0.5cm} & \tau>1
		\end{array} \right.
		\label{eq22}
	\end{eqnarray}
	\begin{eqnarray}
		{\cal F}_{1/2}^{h_1} (\tau,\lambda) & = & \left[I_1(\tau,\lambda) - I_2(\tau,\lambda)
		\right] \\
		{\cal F}_{1}^{h_1} (\tau,\lambda) & = & c_W \left\{ 4\left(3-\frac{s_W^2}{c_W^2} \right)I_2(\tau,\lambda) + \left\{ \left(1+\frac{2}{\tau}\right) \frac{s_W^2}{c_W^2} - \left( 5+\frac{2}{\tau} \right)\right\} I_1(\tau,\lambda)\right\}
		\label{eq24}
	\end{eqnarray}
	with $\hat{v}_f=2I_f^3-4 Q_f s_W^2$. The functions $I_1$ et $I_2$ 
	are given by,
	\begin{eqnarray}
		I_1(\tau,\lambda) &=& \frac{\tau\lambda}{2(\tau-\lambda)}
		+ \frac{\tau^2\lambda^2}{2(\tau-\lambda)^2} \left[ f(\tau^{-1})-f(\lambda^{-1}) 
		\right] + \frac{\tau^2\lambda}{(\tau-\lambda)^2} \left[ g(\tau ^{-1}) - 
		g(\lambda^{-1}) \right]\\
		I_2(\tau,\lambda) &=& - \frac{\tau\lambda}{2(\tau-\lambda)}\left[ f(\tau
		^{-1})- f(\lambda^{-1}) \right]
	\end{eqnarray}.
	the function $f(\tau)$ is defined above in Eq.~(\ref{eq22}) whereas the $g(\tau)$ function can be expressed as,
	\begin{equation}
		g(\tau) = \left\{ \begin{array}{ll}
			\displaystyle \sqrt{\tau^{-1}-1} \arcsin \sqrt{\tau} & \tau \ge 1 \\
			\displaystyle \frac{\sqrt{1-\tau^{-1}}}{2} \left[ \log \frac{1+\sqrt{1-\tau
					^{-1}}}{1-\sqrt{1-\tau^{-1}}} - i\pi \right] & \tau  < 1
		\end{array} \right.
		\label{eq:gtau}
	\end{equation}
   ${\cal F}^{	h_1}_0 (\tau_{i},\lambda_{i})$, $i=H_1^{\pm},H_2^{\pm},H^{\pm\pm}$ factors reflecting the charged contributions
	for $\Gamma_{\gamma\,Z}$ can be read in terms of the function $I_1(\tau,\lambda)$ previously defined as follows :
	\begin{equation}
		\label{fhp_fhpp}
		{\cal F}^{h_1}_0 (\tau_{H_i^{\pm}},\lambda_{H_i^{\pm}}) = 2 I_1(\tau_{H_i^{\pm}},\lambda_{H_i^{\pm}})\ i=1,2 \quad ,\quad {\cal F}^{h_1}_0 (\tau_{H^{\pm\pm}},\lambda_{H^{\pm\pm}}) = 4 I_1(\tau_{H^{\pm\pm}},\lambda_{H^{\pm\pm}}) 
	\end{equation}
	with now $\tau_i= 4M_i^2/M_H^2$, 
	$\lambda_i = 4M_i^2 /M_Z^2$. Where the $\lambda_{ZH^{\pm}H_1^{\mp}}$, $\lambda_{ZH^{\pm}H_2^{\mp}}$,  $\lambda_{ZH^{\pm\pm}H^{\mp\mp}}$ trilinear couplings can be expressed as :
	\begin{eqnarray}
		\lambda_{ZH_1^{\pm}H_1^{\mp}}  & = & [(c_w^2-s_w^2)(C_{21}^2+C_{22}^2)-2 s_w^2 C_{23}^2]/(s_w\,c_w)\\
		\lambda_{ZH_2^{\pm}H_2^{\mp}}  & = & [(c_w^2-s_w^2)(C_{31}^2+C_{32}^2)-2 s_w^2 C_{33}^2]/(s_w\,c_w)\\
		\lambda_{ZH^{\pm\pm}H^{\mp\mp}}& = & 4(c_w^2-s_w^2)/(s_w\,c_w)
		\label{eq:couplagesZHpZHpp}
	\end{eqnarray}	
where $s_w$ and $c_w$  are the sine and cosine of the Weinberg angle ($\theta_w$), respectively.
\vspace{0.5cm}
	\section{The Higgs sector}
The bilinear part of the Higgs potential in Eq.\ref{scalar_pot} is given by:
\begin{eqnarray}
	V^{(2)}_{H_1,H_2,\Delta}&=&\frac{1}{2} \begin{pmatrix} \rho_1,\rho_2, \delta^0 \end{pmatrix} 
	\mathcal{M}_{{\mathcal{CP}}_{even}}^2
	\begin{pmatrix} \rho_1 \\ \rho_2 \\ \rho_0 \end{pmatrix} +
	\frac{1}{2} \begin{pmatrix} \eta_1, \eta_2, \eta_0 \end{pmatrix}
	\mathcal{M}_{{\mathcal{CP}}_{odd}}^2
	\begin{pmatrix} \eta_1 \\ \eta_2 \\ \eta_0 \end{pmatrix} \nonumber\\
	&&+\begin{pmatrix} \phi^-_1,\phi^-_2,\delta^-  \end{pmatrix}
	\mathcal{M}_{\pm}^2
	\begin{pmatrix} \phi^+_1 \\ \phi^+_2\\ \delta^+  \end{pmatrix} +
	\delta^{++}{\mathcal{M}_{\pm\pm}^2}\delta^{--} \cdots, 
\end{eqnarray}
In the subsequent appendices, he elements of these mass matrices are explicitly presented.
	\subsection{ The Scalar sector}
	\label{scalar}
	In the CP-even scalar sector, the mixing of the states ($\rho_{1},\rho_{2},\rho_{0}$) leads to a total of three CP-even physical Higgs bosons ($h_1,h_2,h_3$). The neutral scalar mass matrix reads :

	\begin{eqnarray}
		{\mathcal{M}}_{{\mathcal{CP}}_{even}}^2=\left(
		\begin{array}{ccc}
			m_{\rho_{1}\rho_{1}}^2  &  m_{\rho_{2}\rho_{1}}^2  &  m_{\rho_0\rho_{1}}^2   \\
			m_{\rho_{1}\rho_{2}}^2  &  m_{\rho_{2}\rho_{2}}^2  &  m_{\rho_0\rho_{2}}^2   \\
			m_{\rho_{1}\rho_0}^2  &  m_{\rho_{2}\rho_0}^2  &  m_{\rho_0\rho_0}^2      
		\end{array}
		\right)
		\label{matrix-cp-even}   
	\end{eqnarray}
	Its diagonal terms are,
	\begin{eqnarray}
		m_{\rho_{1}\rho_{1}}^2 &=& \lambda_1 v_1^2+\frac{v_2 \left(\sqrt{2} m_{3}^2 + \mu_3 v_t\right)}{\sqrt{2} v_1} \nonumber\\
		m_{\rho_{2}\rho_{2}}^2 &=& \lambda_2 v_2^2+\frac{v_1 \left(\sqrt{2} m_{3}^2+ \mu_3 v_t\right)}{\sqrt{2} v_2} \nonumber\\
		m_{\rho_0\rho_0}^2 &=& \frac{4 \left(\bar{\lambda}_{8}+\bar{\lambda}_{9}\right) v_t^3+\sqrt{2} \left(\mu_1 v_1^2+\mu_3 v_2 v_1+\mu_2 v_2^2\right)}{2 v_t}
		\label{diago-cp-even}
	\end{eqnarray}
	and the off-diagonal terms are defined by,
	\begin{eqnarray}
		m_{\rho_{2}\rho_{1}}^2 &=& m_{\rho_{1}\rho_{2}}^2 = \frac{1}{\sqrt{2}} \left(\sqrt{2} v_1 v_2 \lambda_{345} - \sqrt{2} m_{3}^2-\mu_3 v_t\right) \nonumber\\
		m_{\rho_0\rho_{1}}^2 &=& m_{\rho_{1}\rho_0}^2 = \frac{1}{\sqrt{2}} \left(\sqrt{2} v_1 v_t (\lambda_6+\lambda_8) - (2 \mu_1 v_1 + \mu_3 v_2)\right) \nonumber\\
		m_{\rho_0\rho_{2}}^2 &=& m_{\rho_{2}\rho_0}^2 = \frac{1}{\sqrt{2}} \left(\sqrt{2} v_2 v_t (\lambda_7+\lambda_9) - (2 \mu_2 v_2 + \mu_3 v_1)\right) 
		\label{off-diago-cp-even} 
	\end{eqnarray}
	The mass matrix can be diagonalized using an orthogonal matrix ${\mathcal{E}}$  parameterized as follows:
	\begin{eqnarray}
		{\mathcal{E}} =\left( \begin{array}{ccc}
			c_{\alpha_1} c_{\alpha_2} & s_{\alpha_1} c_{\alpha_2} & s_{\alpha_2}\\
			-(c_{\alpha_1} s_{\alpha_2} s_{\alpha_3} + s_{\alpha_1} c_{\alpha_3})
			& c_{\alpha_1} c_{\alpha_3} - s_{\alpha_1} s_{\alpha_2} s_{\alpha_3}
			& c_{\alpha_2} s_{\alpha_3} \\
			- c_{\alpha_1} s_{\alpha_2} c_{\alpha_3} + s_{\alpha_1} s_{\alpha_3} &
			-(c_{\alpha_1} s_{\alpha_3} + s_{\alpha_1} s_{\alpha_2} c_{\alpha_3})
			& c_{\alpha_2}  c_{\alpha_3}
		\end{array} \right)
		\label{cpevenmatrix}
	\end{eqnarray}
	where the mixing angles $\alpha_1$, $\alpha_2$ and $\alpha_3$ are in  the range
	\begin{eqnarray}
		- \frac{\pi}{2} \le \alpha_{1,2,3} \le \frac{\pi}{2} \;.
	\end{eqnarray}
	the rotation between the two basis ($\rho_{1},\rho_{2},\rho_{0}$) and ($h_1,h_2,h_3$) diagonalizes the mass matrix ${\mathcal{M}}_{{\mathcal{CP}}_{even}}^2$ as,
	\begin{eqnarray}
		{\mathcal{E}}{\mathcal{M}}_{{{\mathcal{CP}}_{even}}}^2{\mathcal{E}}^T&=&diag(m^2_{h_1},m^2_{h_2},m^2_{h_3})
		\label{rota-matrix-cp-even}
	\end{eqnarray}
	and leads to three mass eigenstates, ordered by ascending mass as:
	\begin{eqnarray}
		m^2_{h_1} < m^2_{h_2} < m^2_{h_3} \;.
	\end{eqnarray}	

	\subsection{The Charged Sector}
	\vspace{0.5cm}
	%--------------------------------------
	\subsubsection*{Mass of the doubly charged bosons}
	%--------------------------------------
	%
	The doubly charged mass $m_{H^{\pm\pm}}^2$, corresponding to the doubly charged eigenstate $\delta^{\pm\pm}$, can simply be determined by collecting all the coefficients of $\delta^{++}\delta^{--}$ in the scalar potential. When that is done the mass reads,
	%    
	\begin{eqnarray}
		m_{H^{\pm\pm}}^2=\frac{\sqrt{2}\mu_1 v_1^2 + \sqrt{2}\mu_3 v_1 v_2 + \sqrt{2}\mu_2 v_2^2 - \lambda_8 v_1^2 v_t
			- \lambda_9 v_2^2 v_t - 2 \bar{\lambda}_9 v_t^3}{2v_t}  \label{eq:mHpmpm}
	\end{eqnarray}
	%--------------------------------------
	\subsubsection*{Mass of the simply charged bosons}
	%--------------------------------------
	%
	
In the Charged scalar sector, the mixing of the states ($\phi_1^\pm,\phi_2^\pm,\delta^\pm$) leads to a total of three charged states, ($G^\pm,H_1^\pm,H_2^\pm$). The rotation between the physical and non-physical states can be represented as : 
\begin{eqnarray}
	\begin{matrix}
		\left(\begin{matrix}
			G^\pm\\
			H_1^\pm\\
			H_2^\pm
		\end{matrix}\right)={\mathcal{C}}\left(\begin{matrix}
			\phi_1^\pm\\
			\phi_2^\pm\\
			\delta^\pm
		\end{matrix}\right)\quad\,
	\end{matrix}
\end{eqnarray}

The mass-squared matrix for the simply charged field in the ($\phi_1^{-},\phi_2^{-},\delta^{-}$) basis reads as:
\begin{eqnarray}
	{\mathcal{M}}_{\pm}^2= \left(
	\begin{array}{ccc}
		{\mathcal{M}}^{\pm}_{11} & \frac{1}{2} \left(\lambda_{45}^+ v_1 v_2-2 m_{3}^2\right) & \frac{1}{4} \left(v_1 A - 2 \mu_3 v_2\right) \\
		\frac{1}{2} \left(\lambda_{45}^+ v_1 v_2-2 m_{3}^2\right) & {\mathcal{M}}^{\pm}_{22} & \frac{1}{4} \left(v_2 B-2 \mu_3 v_1\right) \\
		\frac{1}{4} \left(v_1 A - 2 \mu_3 v_2\right) & \frac{1}{4} \left(v_2 B-2 \mu_3 v_1\right) & {\mathcal{M}}^{\pm}_{33}\\
	\end{array}
	\right)
	\label{matrix_charged}
\end{eqnarray}
where $A=\sqrt{2} \lambda_8 v_t-4\mu_1$, $B=\sqrt{2} \lambda_9 v_t-4\mu_2$ while the diagonal terms read as,
\begin{eqnarray}
	{\mathcal{M}}^{\pm}_{11} &=& \frac{2 m_{3}^2 v_2+v_1 v_t \left(2 \sqrt{2} \mu_1-\lambda_8 v_t\right)+\sqrt{2} \mu_3 v_2 v_t- \lambda_{45}^+ v_1 v_2^2}{2 v_1} \nonumber\\
	{\mathcal{M}}^{\pm}_{22} &=& \frac{2 m_{3}^2 v_1+v_2 v_t \left(2\sqrt{2} \mu_2-\lambda_9 v_t \right)+\sqrt{2} \mu_3 v_1 v_t - \lambda_{45}^+ v_1^2 v_2}{2 v_2} \nonumber\\
	{\mathcal{M}}^{\pm}_{33} &=& \frac{v_1^2 \left(2\sqrt{2} \mu_1-\lambda_8 v_t\right)+v_2^2 \left(2\sqrt{2} \mu_2-\lambda_9 v_t\right)+2 \sqrt{2} \mu_3 v_2 v_1-2\bar{\lambda_9}v_t^3}{4 v_t}  
	\label{diago-charged}
\end{eqnarray}
Among the three eigenvalues of the matrix ${\mathcal{M}}_{\pm}^2$, one is zero and corresponds to the charged Goldstone boson $G^\pm$, The remaining two eigenvalues correspond to the singly charged Higgs bosons, denoted as $m_{H_{1}^\pm}^2$ and $m_{H_{2}^\pm}^2$ and are given by,
%\textcolor{black}{
	\begin{eqnarray}
		m^2_{H^\pm_{1,2}}=\frac{1}{4 v_0^2 v_t}\Big[-v_0 \left(v_0 \left(2 {\mathcal{M}}_{12}^\pm \text{cs}_{\beta } \text{se}_{\beta } v_t+\kappa \right)+2 \sqrt{2} v_t^2 \left({\mathcal{M}}_{23}^\pm \text{cs}_{\beta }+{\mathcal{M}}_{13}^\pm \text{se}_{\beta }\right)\right)\mp\text{cs}_{\beta } \text{se}_{\beta }\sqrt{ \mathcal{Y}}\Big]
	\end{eqnarray}
	where $\text{c}_x$, $\text{s}_x$, $\text{cs}_x$, $\text{se}_x$ stand for the $\cos(x)$, $\sin(x)$, $\csc(x)$, $\sec(x)$ respectively, while $v_0=\sqrt{v_1^2+v_2^2}$, $v=\sqrt{v_1^2+v_2^2+2 v_t^2}$, $\kappa=\sqrt{2} v_0 \left({\mathcal{M}}_{13}^\pm c_{\beta }+{\mathcal{M}}_{23}^\pm s_{\beta }\right)$ and
	\begin{eqnarray}
		\mathcal{Y}&=&v_0^2 \Big(\big(v_0 \left(\kappa  c_{\beta } s_{\beta }+2 {\mathcal{M}}^{\pm}_{12} v_t \right)+2 \sqrt{2} v_t^2 \left({\mathcal{M}}^{\pm}_{23} c_{\beta }+{\mathcal{M}}^{\pm}_{13} s_{\beta }\right)\big)^2\nonumber\\
		&-&4 v^2 s_{2 \beta } v_t \big(\kappa  {\mathcal{M}}^{\pm}_{12}+2 {\mathcal{M}}^{\pm}_{13} {\mathcal{M}}^{\pm}_{23} v_t\big)\Big)
	\end{eqnarray}

The above symmetric squared matrix ${\mathcal{M}}_{\pm}^2$ can be diagonalized via ${\mathcal{C}}$ as follows,
%.
\begin{eqnarray}
	{\mathcal{C}}{\mathcal{M}}_{\pm}^2{\mathcal{C}}^T&=&diag(m^2_{G^\pm},m^2_{H_1^\pm},m^2_{H_2^\pm})
	\label{rota-matrix-charged}
\end{eqnarray}
where the ${\mathcal{C}}$ rotation matrix is described by three mixing angles $\theta^\pm_1$, $\theta^\pm_2$ and $\theta^\pm_3$, and the corresponding expressions
for the ${\mathcal{C}}$ elements as a function of the input parameters of our model are given by,

\begin{eqnarray}
	{\mathcal{C}} =\left( \begin{array}{ccc}
		{\mathcal{C}}_{11} & {\mathcal{C}}_{12} & {\mathcal{C}}_{13}\\
		{\mathcal{C}}_{21} & {\mathcal{C}}_{22} & {\mathcal{C}}_{23} \\
		{\mathcal{C}}_{31} & {\mathcal{C}}_{32} & {\mathcal{C}}_{33}
	\end{array} \right)=\left( \begin{array}{ccc}
		c_{\theta^\pm_1} c_{\theta^\pm_2} & s_{\theta^\pm_1} c_{\theta^\pm_2} & s_{\theta^\pm_2}\\
		-(c_{\theta^\pm_1} s_{\theta^\pm_2} s_{\theta^\pm_3} + s_{\theta^\pm_1} c_{\theta^\pm_3})
		& c_{\theta^\pm_1} c_{\theta^\pm_3} - s_{\theta^\pm_1} s_{\theta^\pm_2} s_{\theta^\pm_3}
		& c_{\theta^\pm_2} s_{\theta^\pm_3} \\
		- c_{\theta^\pm_1} s_{\theta^\pm_2} c_{\theta^\pm_3} + s_{\theta^\pm_1} s_{\theta^\pm_3} &
		-(c_{\theta^\pm_1} s_{\theta^\pm_3} + s_{\theta^\pm_1} s_{\theta^\pm_2} c_{\theta^\pm_3})
		& c_{\theta^\pm_2}  c_{\theta^\pm_3}
	\end{array} \right)
	\label{eq:mixingmatrix}
\end{eqnarray}

\begin{eqnarray}
	&& {\mathcal{C}}_{11}=\frac{v_1 }{v},\hspace{3.6cm}{\mathcal{C}}_{12}=\frac{v_2}{v},\hspace{3.6cm}
	{\mathcal{C}}_{13}=\sqrt{2}\frac{v_t}{v} \\
	&& {\mathcal{C}}_{21}=\frac{x_1}{\sqrt{\mathcal{N}}},\hspace{3.25cm}{\mathcal{C}}_{22}=\frac{x_2}{\sqrt{\mathcal{N}}},\hspace{3.2cm}{\mathcal{C}}_{22}=\frac{1}{\sqrt{\mathcal{N}}}\\
	&& {\mathcal{C}}_{31}={\mathcal{C}}_{21}[m^2_{H_1^\pm}\to m^2_{H_2^\pm}],
	\hspace{1cm}
	{\mathcal{C}}_{32}={\mathcal{C}}_{32}[m^2_{H_1^\pm}\to m^2_{H_2^\pm}],
	\hspace{1cm}
	{\mathcal{C}}_{33}={\mathcal{C}}_{23}[m^2_{H_1^\pm}\to m^2_{H_2^\pm}]
\end{eqnarray}
where $v_0=\sqrt{v_1^2+v_2^2}$ and
% {\mathcal{M}}^{\pm}
{\begin{eqnarray}
		x_1=\frac{v_0 c_{\beta } \left(v_0 \left({\mathcal{M}}^{\pm}_{12} \left({\mathcal{M}}^{\pm}_{13} c_{\beta }+{\mathcal{M}}^{\pm}_{23} s_{\beta }\right)+{\mathcal{M}}^{\pm}_{13} m_{H^\pm_1}^2 s_{\beta }\right)+\sqrt{2} {\mathcal{M}}^{\pm}_{13} {\mathcal{M}}^{\pm}_{23} v_t\right)}{\sqrt{2} v_0 v_t \left(m_{H^\pm_1}^2 \left(M_{23} c_{\beta }+{\mathcal{M}}^{\pm}_{13} s_{\beta }\right)+{\mathcal{M}}^{\pm}_{12} \left({\mathcal{M}}^{\pm}_{13} c_{\beta }+{\mathcal{M}}^{\pm}_{23} s_{\beta }\right)\right)+v_0^2 m_{H^\pm_1}^2 \left(c_{\beta } m_{H^\pm_1}^2 s_{\beta }+{\mathcal{M}}^{\pm}_{12}\right)+2 {\mathcal{M}}^{\pm}_{13} {\mathcal{M}}^{\pm}_{23} v_t^2}\nonumber\\
\end{eqnarray}
{\begin{eqnarray}
		x_2=\frac{v_0 s_{\beta } \left({\mathcal{M}}^{\pm}_{23} \left(v_0 c_{\beta } m_{H^\pm_1}^2+\sqrt{2} {\mathcal{M}}^{\pm}_{13} v_t\right)+{\mathcal{M}}^{\pm}_{12} v_0 \left({\mathcal{M}}^{\pm}_{13} c_{\beta }+{\mathcal{M}}^{\pm}_{23} s_{\beta }\right)\right)}{\sqrt{2} v_0 v_t \left(m_{H^\pm_1}^2 \left({\mathcal{M}}^{\pm}_{23} c_{\beta }+{\mathcal{M}}^{\pm}_{13} s_{\beta }\right)+{\mathcal{M}}^{\pm}_{12} \left({\mathcal{M}}^{\pm}_{13} c_{\beta }+{\mathcal{M}}^{\pm}_{23} s_{\beta }\right)\right)+v_0^2 m_{H^\pm_1}^2 \left(c_{\beta } m_{H^\pm_1}^2 s_{\beta }+{\mathcal{M}}^{\pm}_{12}\right)+2 {\mathcal{M}}^{\pm}_{13} {\mathcal{M}}^{\pm}_{23} v_t^2}\nonumber\\
\end{eqnarray}}
\begin{eqnarray}
	{\mathcal{N}}=\sqrt{1+x_1^2+x_2^2}
\end{eqnarray}}
For the input parameters implemented in $2HDMcT$, we used the following hybrid parametrization,
	\begin{eqnarray}
		\mathcal{P}_I = \left\{\alpha_1,\alpha_2,\alpha_3,m_{h_1},m_{h_2},m_{h_3},m_{H^{\pm\pm}}, \lambda_{1},\lambda_{3},\lambda_{4},\lambda_{6},\lambda_{8},\bar{\lambda}_{8},\bar{\lambda}_{9},\mu_1,v_t,\tan\beta \right\}
		\label{eq:para1}
	\end{eqnarray}
	with $\tan\beta=v_2/v_1$. One can easily express the set of Lagrangian parameters in Eq.\ref{scalar_pot} in terms of those given by. \ref{eq:para1}.
	\begin{eqnarray}
		&& \lambda_2 = \frac{-\mathcal{B} c_{\beta }^2+\lambda_1 v_0^2 c_{\beta }^4+\left({\mathcal{E}}_{12}^2 m_{h_1}^2+{\mathcal{E}}_{22}^2 m_{h_2}^2+{\mathcal{E}}_{32}^2 m_{h_3}^2\right)s_{\beta }^2}{v_0^2 s_{\beta }^4}\nonumber\\
		&& \lambda_5 =\frac{\left(\mathcal{B} - \lambda^+_{34} v_0^2 s_{\beta }^2\right)c_{\beta }-\lambda_1v_0^2c_{\beta }^3+\left({\mathcal{E}}_{11} {\mathcal{E}}_{12} m_{h_1}^2+{\mathcal{E}}_{21} {\mathcal{E}}_{22} m_{h_2}^2+{\mathcal{E}}_{31} {\mathcal{E}}_{32} m_{h_3}^2\right)s_{\beta }}{v_0^2 c_{\beta } s_{\beta }^2}\nonumber\\
		&& \lambda_9 = \frac{-\lambda_8 v_0^2 c_{\beta }^2+2 \mathcal{F} - 2 m_{H^{\pm\pm}}^2-2 \left(2 \bar{\lambda}_8+3 \bar{\lambda}_9\right) v_t^2}{v_0^2 s_{\beta }^2}\nonumber\\
		&& \lambda_7 =\frac{v_0 \left(\mathcal{A} c_{\beta }+\mathcal{M} s_{\beta }\right)+v_t \left(-\lambda_6 v_0^2 c_{\beta }^2-2 \mathcal{F}+2 m_{H^{\pm\pm}}^2+2 \left(2 \bar{\lambda}_8+3 \bar{\lambda}_9 \right) v_t^2\right)}{v_0^2 s_{\beta }^2 v_t}\nonumber\\
		&& \mu_2 = \frac{c_{\beta } \left(2 \mathcal{A}+v_0 c_{\beta } \left(\sqrt{2} \mu _1-2 \lambda_{68}^+ v_t\right)\right)}{\sqrt{2} v_0 s_{\beta }^2}\nonumber\\
		&& \mu_3 =\frac{\sqrt{2} \left(-\mathcal{A} v_0 c_{\beta }+v_0^2 c_{\beta }^2 \left(\lambda_{68}^+ v_t-\sqrt{2} \mu _1\right)+v_t \left({\mathcal{E}}_{13}^2 m_{h_1}^2+{\mathcal{E}}_{23}^2 m_{h_2}^2+{\mathcal{E}}_{33}^2 m_{h_3}^2-2 \bar{\lambda}_{89}^+ v_t^2\right)\right)}{v_0^2 c_{\beta } s_{\beta }}\nonumber\\
		&& m_{3}^2 = \frac{\mathcal{A} v_t+v_0 c_{\beta } \left(\mathcal{B}-v_t \left(\lambda_{68}^+ v_t-\sqrt{2} \mu _1\right)\right)-\lambda_1 v_0^3 c_{\beta }^3}{v_0 s_{\beta }}
	\end{eqnarray}
	where $v_0=\sqrt{v_1^2+v_2^2}$ and
	\begin{eqnarray}
		&& \mathcal{A}={\mathcal{E}}_{11} {\mathcal{E}}_{13} m_{h_1}^2+{\mathcal{E}}_{21} {\mathcal{E}}_{23} m_{h_2}^2+{\mathcal{E}}_{31} {\mathcal{E}}_{33} m_{h_3}^2\nonumber\\
		&& \mathcal{B} = {\mathcal{E}}_{11}^2 m_{h_1}^2+{\mathcal{E}}_{21}^2 m_{h_2}^2+{\mathcal{E}}_{31}^2 m_{h_3}^2\nonumber\\
		&& \mathcal{F} = {\mathcal{E}}_{13}^2 m_{h_1}^2+{\mathcal{E}}_{23}^2 m_{h_2}^2+{\mathcal{E}}_{33}^2 m_{h_3}^2\nonumber\\
		&& \mathcal{M} = {\mathcal{E}}_{12} {\mathcal{E}}_{13} m_{h_1}^2+{\mathcal{E}}_{22} {\mathcal{E}}_{23} m_{h_2}^2+{\mathcal{E}}_{32} {\mathcal{E}}_{33} m_{h_3}^2
	\end{eqnarray}

       \bibliographystyle{JHEP}
       \bibliography{references}

\providecommand{\href}[2]{#2}\begingroup\raggedright\begin{thebibliography}{10}

\bibitem{ATLAS:2012yve}
{\bf ATLAS} Collaboration, G.~Aad et~al., {\it {Observation of a new particle
  in the search for the Standard Model Higgs boson with the ATLAS detector at
  the LHC}},  {\em Phys. Lett. B} {\bf 716} (2012) 1--29,
  [\href{http://arxiv.org/abs/1207.7214}{{\tt arXiv:1207.7214}}].

\bibitem{CMS:2012qbp}
{\bf CMS} Collaboration, S.~Chatrchyan et~al., {\it {Observation of a New Boson
  at a Mass of 125 GeV with the CMS Experiment at the LHC}},  {\em Phys. Lett.
  B} {\bf 716} (2012) 30--61, [\href{http://arxiv.org/abs/1207.7235}{{\tt
  arXiv:1207.7235}}].

\bibitem{CMS:2022dwd}
{\bf CMS} Collaboration, A.~Tumasyan et~al., {\it {A portrait of the Higgs
  boson by the CMS experiment ten years after the discovery.}},  {\em Nature}
  {\bf 607} (2022), no.~7917 60--68,
  [\href{http://arxiv.org/abs/2207.00043}{{\tt arXiv:2207.00043}}].

\bibitem{ATLAS:2022vkf}
{\bf ATLAS} Collaboration, G.~Aad et~al., {\it {A detailed map of Higgs boson
  interactions by the ATLAS experiment ten years after the discovery}},  {\em
  Nature} {\bf 607} (2022), no.~7917 52--59,
  [\href{http://arxiv.org/abs/2207.00092}{{\tt arXiv:2207.00092}}]. [Erratum:
  Nature 612, E24 (2022)].

\bibitem{Elmetenawee:2024dxc}
W.~Elmetenawee, {\it {Summary of CMS Higgs Physics}},
  \href{http://arxiv.org/abs/2401.07650}{{\tt arXiv:2401.07650}}.

\bibitem{Zwicky:1933gu}
F.~Zwicky, {\it {Die Rotverschiebung von extragalaktischen Nebeln}},  {\em
  Helv. Phys. Acta} {\bf 6} (1933) 110--127.

\bibitem{Rubin:1970zza}
V.~C. Rubin and W.~K. Ford, Jr., {\it {Rotation of the Andromeda Nebula from a
  Spectroscopic Survey of Emission Regions}},  {\em Astrophys. J.} {\bf 159}
  (1970) 379--403.

\bibitem{Veltman:1980mj}
M.~J.~G. Veltman, {\it {The Infrared - Ultraviolet Connection}},  {\em Acta
  Phys. Polon. B} {\bf 12} (1981) 437.

\bibitem{RevModPhys.88.030501}
T.~Kajita, {\it Nobel lecture: Discovery of atmospheric neutrino oscillations},
   {\em Rev. Mod. Phys.} {\bf 88} (Jul, 2016) 030501.

\bibitem{RevModPhys.88.030502}
A.~B. McDonald, {\it Nobel lecture: The sudbury neutrino observatory:
  Observation of flavor change for solar neutrinos},  {\em Rev. Mod. Phys.}
  {\bf 88} (Jul, 2016) 030502.

\bibitem{Deshpande:1977rw}
N.~G. Deshpande and E.~Ma, {\it {Pattern of Symmetry Breaking with Two Higgs
  Doublets}},  {\em Phys. Rev. D} {\bf 18} (1978) 2574.

\bibitem{PhysRevD.15.1958}
S.~L. Glashow and S.~Weinberg, {\it Natural conservation laws for neutral
  currents},  {\em Phys. Rev. D} {\bf 15} (Apr, 1977) 1958--1965.

\bibitem{Branco:2011iw}
G.~C. Branco, P.~M. Ferreira, L.~Lavoura, M.~N. Rebelo, M.~Sher, and J.~P.
  Silva, {\it {Theory and phenomenology of two-Higgs-doublet models}},  {\em
  Phys. Rept.} {\bf 516} (2012) 1--102,
  [\href{http://arxiv.org/abs/1106.0034}{{\tt arXiv:1106.0034}}].

\bibitem{Dawson:2018dcd}
S.~Dawson, C.~Englert, and T.~Plehn, {\it {Higgs Physics: It ain't over till
  it's over}},  {\em Phys. Rept.} {\bf 816} (2019) 1--85,
  [\href{http://arxiv.org/abs/1808.01324}{{\tt arXiv:1808.01324}}].

\bibitem{Ivanov:2017dad}
I.~P. Ivanov, {\it {Building and testing models with extended Higgs sectors}},
  {\em Prog. Part. Nucl. Phys.} {\bf 95} (2017) 160--208,
  [\href{http://arxiv.org/abs/1702.03776}{{\tt arXiv:1702.03776}}].

\bibitem{PhysRevLett.43.1566}
S.~Weinberg, {\it Baryon- and lepton-nonconserving processes},  {\em Phys. Rev.
  Lett.} {\bf 43} (Nov, 1979) 1566--1570.

\bibitem{Ouazghour:2018mld}
B.~Ait~Ouazghour, A.~Arhrib, R.~Benbrik, M.~Chabab, and L.~Rahili, {\it {Theory
  and phenomenology of a two-Higgs-doublet type-II seesaw model at the LHC run
  2}},  {\em Phys. Rev. D} {\bf 100} (2019), no.~3 035031,
  [\href{http://arxiv.org/abs/1812.07719}{{\tt arXiv:1812.07719}}].

\bibitem{Ouazghour:2023eqr}
B.~Ait~Ouazghour and M.~Chabab, {\it {The two Higgs doublet type-II seesaw
  model: Naturalness and
  B\textasciimacron{}\textrightarrow{}Xs\ensuremath{\gamma} versus heavy Higgs
  masses}},  {\em Phys. Lett. B} {\bf 846} (2023) 138241,
  [\href{http://arxiv.org/abs/2305.08030}{{\tt arXiv:2305.08030}}].

\bibitem{ATLAS:2017eiz}
{\bf ATLAS} Collaboration, M.~Aaboud et~al., {\it {Search for additional heavy
  neutral Higgs and gauge bosons in the ditau final state produced in 36
  fb$^{-1}$ of pp collisions at $ \sqrt{s}=13 $ TeV with the ATLAS detector}},
  {\em JHEP} {\bf 01} (2018) 055, [\href{http://arxiv.org/abs/1709.07242}{{\tt
  arXiv:1709.07242}}].

\bibitem{ATLAS:2017ayi}
{\bf ATLAS} Collaboration, M.~Aaboud et~al., {\it {Search for new phenomena in
  high-mass diphoton final states using 37 fb$^{-1}$ of proton--proton
  collisions collected at $\sqrt{s}=13$ TeV with the ATLAS detector}},  {\em
  Phys. Lett. B} {\bf 775} (2017) 105--125,
  [\href{http://arxiv.org/abs/1707.04147}{{\tt arXiv:1707.04147}}].

\bibitem{CMS:2018amk}
{\bf CMS} Collaboration, A.~M. Sirunyan et~al., {\it {Search for a new scalar
  resonance decaying to a pair of Z bosons in proton-proton collisions at
  $\sqrt{s}=13 $ TeV}},  {\em JHEP} {\bf 06} (2018) 127,
  [\href{http://arxiv.org/abs/1804.01939}{{\tt arXiv:1804.01939}}]. [Erratum:
  JHEP 03, 128 (2019)].

\bibitem{ATLAS:2018oht}
{\bf ATLAS} Collaboration, M.~Aaboud et~al., {\it {Search for a heavy Higgs
  boson decaying into a $Z$ boson and another heavy Higgs boson in the
  $\ell\ell bb$ final state in $pp$ collisions at $\sqrt{s}=13$ TeV with the
  ATLAS detector}},  {\em Phys. Lett. B} {\bf 783} (2018) 392--414,
  [\href{http://arxiv.org/abs/1804.01126}{{\tt arXiv:1804.01126}}].

\bibitem{CMS:2019pzc}
{\bf CMS} Collaboration, A.~M. Sirunyan et~al., {\it {Search for heavy Higgs
  bosons decaying to a top quark pair in proton-proton collisions at $\sqrt{s}
  =$ 13 TeV}},  {\em JHEP} {\bf 04} (2020) 171,
  [\href{http://arxiv.org/abs/1908.01115}{{\tt arXiv:1908.01115}}]. [Erratum:
  JHEP 03, 187 (2022)].

\bibitem{CMS:2019mij}
{\bf CMS} Collaboration, A.~M. Sirunyan et~al., {\it {Search for MSSM Higgs
  bosons decaying to \ensuremath{\mu} + \ensuremath{\mu} \ensuremath{-} in
  proton-proton collisions at s=13TeV}},  {\em Phys. Lett. B} {\bf 798} (2019)
  134992, [\href{http://arxiv.org/abs/1907.03152}{{\tt arXiv:1907.03152}}].

\bibitem{2022arXiv220701046C}
{CMS Collaboration}, {\it {Search for a charged Higgs boson decaying into a
  heavy neutral Higgs boson and a W boson in proton-proton collisions at
  $\sqrt{s}$ = 13 TeV}},  {\em arXiv e-prints} (July, 2022) arXiv:2207.01046,
  [\href{http://arxiv.org/abs/2207.01046}{{\tt arXiv:2207.01046}}].

\bibitem{Gupta:2013zza}
R.~S. Gupta, H.~Rzehak, and J.~D. Wells, {\it {How well do we need to measure
  the Higgs boson mass and self-coupling?}},  {\em Phys. Rev. D} {\bf 88}
  (2013) 055024, [\href{http://arxiv.org/abs/1305.6397}{{\tt
  arXiv:1305.6397}}].

\bibitem{Baglio:2016bop}
J.~Baglio and C.~Weiland, {\it {The triple Higgs coupling: A new probe of
  low-scale seesaw models}},  {\em JHEP} {\bf 04} (2017) 038,
  [\href{http://arxiv.org/abs/1612.06403}{{\tt arXiv:1612.06403}}].

\bibitem{Tian:2016qlk}
{\bf ILC physics, detector study} Collaboration, J.~Tian and K.~Fujii, {\it
  {Measurement of Higgs boson couplings at the International Linear Collider}},
   {\em Nucl. Part. Phys. Proc.} {\bf 273-275} (2016) 826--833.

\bibitem{Durig:2016jrs}
C.~F. D\"urig, {\em {Measuring the Higgs Self-coupling at the International
  Linear Collider}}.
\newblock PhD thesis, Hamburg U., Hamburg, 2016.

\bibitem{Liu:2018peg}
T.~Liu, K.-F. Lyu, J.~Ren, and H.~X. Zhu, {\it {Probing the quartic Higgs boson
  self-interaction}},  {\em Phys. Rev. D} {\bf 98} (2018), no.~9 093004,
  [\href{http://arxiv.org/abs/1803.04359}{{\tt arXiv:1803.04359}}].

\bibitem{Moortgat-Pick:2015lbx}
A.~Arbey et~al., {\it {Physics at the e+ e- Linear Collider}},  {\em Eur. Phys.
  J. C} {\bf 75} (2015), no.~8 371,
  [\href{http://arxiv.org/abs/1504.01726}{{\tt arXiv:1504.01726}}].

\bibitem{ILCInternationalDevelopmentTeam:2022izu}
{\bf ILC International Development Team} Collaboration, A.~Aryshev et~al., {\it
  {The International Linear Collider: Report to Snowmass 2021}},
  \href{http://arxiv.org/abs/2203.07622}{{\tt arXiv:2203.07622}}.

\bibitem{Bambade:2019fyw}
P.~Bambade et~al., {\it {The International Linear Collider: A Global Project}},
   \href{http://arxiv.org/abs/1903.01629}{{\tt arXiv:1903.01629}}.

\bibitem{CLICPhysicsWorkingGroup:2004qvu}
{\bf CLIC Physics Working Group} Collaboration, E.~Accomando et~al., {\it
  {Physics at the CLIC multi-TeV linear collider}},  in {\em {11th
  International Conference on Hadron Spectroscopy}}, CERN Yellow Reports:
  Monographs, 6, 2004.
\newblock \href{http://arxiv.org/abs/hep-ph/0412251}{{\tt hep-ph/0412251}}.

\bibitem{CEPCStudyGroup:2018rmc}
{\bf CEPC Study Group} Collaboration, {\it {CEPC Conceptual Design Report:
  Volume 1 - Accelerator}},  \href{http://arxiv.org/abs/1809.00285}{{\tt
  arXiv:1809.00285}}.

\bibitem{CEPCStudyGroup:2018ghi}
{\bf CEPC Study Group} Collaboration, M.~Dong et~al., {\it {CEPC Conceptual
  Design Report: Volume 2 - Physics \& Detector}},
  \href{http://arxiv.org/abs/1811.10545}{{\tt arXiv:1811.10545}}.

\bibitem{CEPCStudyGroup:2023quu}
{\bf CEPC Study Group} Collaboration, W.~Abdallah et~al., {\it {CEPC Technical
  Design Report -- Accelerator}},  \href{http://arxiv.org/abs/2312.14363}{{\tt
  arXiv:2312.14363}}.

\bibitem{TLEPDesignStudyWorkingGroup:2013myl}
{\bf TLEP Design Study Working Group} Collaboration, M.~Bicer et~al., {\it
  {First Look at the Physics Case of TLEP}},  {\em JHEP} {\bf 01} (2014) 164,
  [\href{http://arxiv.org/abs/1308.6176}{{\tt arXiv:1308.6176}}].

\bibitem{FCC:2018byv}
{\bf FCC} Collaboration, A.~Abada et~al., {\it {FCC Physics Opportunities}:
  {Future Circular Collider Conceptual Design Report Volume 1}},  {\em Eur.
  Phys. J. C} {\bf 79} (2019), no.~6 474.

\bibitem{LCCPhysicsWorkingGroup:2019fvj}
{\bf LCC Physics Working Group} Collaboration, K.~Fujii et~al., {\it {Tests of
  the Standard Model at the International Linear Collider}},
  \href{http://arxiv.org/abs/1908.11299}{{\tt arXiv:1908.11299}}.

\bibitem{PhysRevD.52.3919}
A.~Abbasabadi, D.~Bowser-Chao, D.~A. Dicus, and W.~W. Repko, {\it
  Higgs-boson--photon associated production at e\ifmmode \bar{e}\else
  \={e}\fi{} colliders},  {\em Phys. Rev. D} {\bf 52} (Oct, 1995) 3919--3928.

\bibitem{1997NuPhB.491...68D}
A.~{Djouadi}, V.~{Driesen}, W.~{Hollik}, and J.~{Rosiek}, {\it {Associated
  production of Higgs bosons and a photon in high-energy e$^{+}$e$^{-}$
  collisions}},  {\em Nuclear Physics B} {\bf 491} (Feb., 1997) 68--102,
  [\href{http://arxiv.org/abs/hep-ph/9609420}{{\tt hep-ph/9609420}}].

\bibitem{Barroso:1985et}
A.~Barroso, J.~Pulido, and J.~C. Romao, {\it {HIGGS PRODUCTION AT e+ e-
  COLLIDERS}},  {\em Nucl. Phys. B} {\bf 267} (1986) 509--530.

\bibitem{Arhrib:2014pva}
A.~Arhrib, R.~Benbrik, and T.-C. Yuan, {\it {Associated Production of Higgs at
  Linear Collider in the Inert Higgs Doublet Model}},  {\em Eur. Phys. J. C}
  {\bf 74} (2014) 2892, [\href{http://arxiv.org/abs/1401.6698}{{\tt
  arXiv:1401.6698}}].

\bibitem{Rahili:2019ixf}
L.~Rahili, A.~Arhrib, and R.~Benbrik, {\it {Associated production of SM Higgs
  with a photon in type-II seesaw models at the ILC}},  {\em Eur. Phys. J. C}
  {\bf 79} (2019), no.~11 940, [\href{http://arxiv.org/abs/1909.07793}{{\tt
  arXiv:1909.07793}}].

\bibitem{2016EPJC}
S.~{Heinemeyer} and C.~{Schappacher}, {\it {Neutral Higgs boson production at
  e\^+e\^- colliders in the complex MSSM: a full one-loop analysis}},  {\em
  European Physical Journal C} {\bf 76} (Apr., 2016) 220,
  [\href{http://arxiv.org/abs/1511.06002}{{\tt arXiv:1511.06002}}].

\bibitem{Demirci:2019ush}
M.~Demirci, {\it {Associated production of Higgs boson with a photon at
  electron-positron colliders}},  {\em Phys. Rev. D} {\bf 100} (2019), no.~7
  075006, [\href{http://arxiv.org/abs/1905.09363}{{\tt arXiv:1905.09363}}].

\bibitem{Aoki:2022dxg}
{\bf ILD concept group} Collaboration, Y.~Aoki, K.~Fujii, and J.~Tian, {\it
  {Study of $e^+e^- \to \gamma h$ at the ILC}},
  \href{http://arxiv.org/abs/2203.07202}{{\tt arXiv:2203.07202}}.

\bibitem{Chen:2014lla}
C.-H. Chen and T.~Nomura, {\it {Inert Dark Matter in Type-II Seesaw}},  {\em
  JHEP} {\bf 09} (2014) 120, [\href{http://arxiv.org/abs/1404.2996}{{\tt
  arXiv:1404.2996}}].

\bibitem{FileviezPerez:2008jbu}
P.~Fileviez~Perez, T.~Han, G.-y. Huang, T.~Li, and K.~Wang, {\it {Neutrino
  Masses and the CERN LHC: Testing Type II Seesaw}},  {\em Phys. Rev. D} {\bf
  78} (2008) 015018, [\href{http://arxiv.org/abs/0805.3536}{{\tt
  arXiv:0805.3536}}].

\bibitem{Cai:2017mow}
Y.~Cai, T.~Han, T.~Li, and R.~Ruiz, {\it {Lepton Number Violation: Seesaw
  Models and Their Collider Tests}},  {\em Front. in Phys.} {\bf 6} (2018) 40,
  [\href{http://arxiv.org/abs/1711.02180}{{\tt arXiv:1711.02180}}].

\bibitem{King:2015aea}
S.~F. King, {\it {Models of Neutrino Mass, Mixing and CP Violation}},  {\em J.
  Phys. G} {\bf 42} (2015) 123001, [\href{http://arxiv.org/abs/1510.02091}{{\tt
  arXiv:1510.02091}}].

\bibitem{Ashanujjaman:2021txz}
S.~Ashanujjaman and K.~Ghosh, {\it {Revisiting type-II see-saw: present limits
  and future prospects at LHC}},  {\em JHEP} {\bf 03} (2022) 195,
  [\href{http://arxiv.org/abs/2108.10952}{{\tt arXiv:2108.10952}}].

\bibitem{Ramsey-Musolf:2019lsf}
M.~J. Ramsey-Musolf, {\it {The electroweak phase transition: a collider
  target}},  {\em JHEP} {\bf 09} (2020) 179,
  [\href{http://arxiv.org/abs/1912.07189}{{\tt arXiv:1912.07189}}].

\bibitem{Peskin:1991sw}
M.~E. Peskin and T.~Takeuchi, {\it {Estimation of oblique electroweak
  corrections}},  {\em Phys. Rev.} {\bf D46} (1992) 381--409.

\bibitem{Grimus:2008nb}
W.~Grimus, L.~Lavoura, O.~M. Ogreid, and P.~Osland, {\it {The Oblique
  parameters in multi-Higgs-doublet models}},  {\em Nucl. Phys. B} {\bf 801}
  (2008) 81--96, [\href{http://arxiv.org/abs/0802.4353}{{\tt
  arXiv:0802.4353}}].

\bibitem{ParticleDataGroup:2022pth}
{\bf Particle Data Group} Collaboration, R.~L. Workman et~al., {\it {Review of
  Particle Physics}},  {\em PTEP} {\bf 2022} (2022) 083C01.

\bibitem{Bahl:2022igd}
H.~Bahl, T.~Biek\"otter, S.~Heinemeyer, C.~Li, S.~Paasch, G.~Weiglein, and
  J.~Wittbrodt, {\it {HiggsTools: BSM scalar phenomenology with new versions of
  HiggsBounds and HiggsSignals}},  {\em Comput. Phys. Commun.} {\bf 291} (2023)
  108803, [\href{http://arxiv.org/abs/2210.09332}{{\tt arXiv:2210.09332}}].

\bibitem{Bechtle:2013xfa}
P.~Bechtle, S.~Heinemeyer, O.~St\r{a}l, T.~Stefaniak, and G.~Weiglein, {\it
  {$HiggsSignals$: Confronting arbitrary Higgs sectors with measurements at the
  Tevatron and the LHC}},  {\em Eur. Phys. J. C} {\bf 74} (2014), no.~2 2711,
  [\href{http://arxiv.org/abs/1305.1933}{{\tt arXiv:1305.1933}}].

\bibitem{Bechtle:2014ewa}
P.~Bechtle, S.~Heinemeyer, O.~St\r{a}l, T.~Stefaniak, and G.~Weiglein, {\it
  {Probing the Standard Model with Higgs signal rates from the Tevatron, the
  LHC and a future ILC}},  {\em JHEP} {\bf 11} (2014) 039,
  [\href{http://arxiv.org/abs/1403.1582}{{\tt arXiv:1403.1582}}].

\bibitem{Bechtle:2020uwn}
P.~Bechtle, S.~Heinemeyer, T.~Klingl, T.~Stefaniak, G.~Weiglein, and
  J.~Wittbrodt, {\it {HiggsSignals-2: Probing new physics with precision Higgs
  measurements in the LHC 13 TeV era}},  {\em Eur. Phys. J. C} {\bf 81} (2021),
  no.~2 145, [\href{http://arxiv.org/abs/2012.09197}{{\tt arXiv:2012.09197}}].

\bibitem{Bechtle:2008jh}
P.~Bechtle, O.~Brein, S.~Heinemeyer, G.~Weiglein, and K.~E. Williams, {\it
  {HiggsBounds: Confronting Arbitrary Higgs Sectors with Exclusion Bounds from
  LEP and the Tevatron}},  {\em Comput. Phys. Commun.} {\bf 181} (2010)
  138--167, [\href{http://arxiv.org/abs/0811.4169}{{\tt arXiv:0811.4169}}].

\bibitem{Bechtle:2011sb}
P.~Bechtle, O.~Brein, S.~Heinemeyer, G.~Weiglein, and K.~E. Williams, {\it
  {HiggsBounds 2.0.0: Confronting Neutral and Charged Higgs Sector Predictions
  with Exclusion Bounds from LEP and the Tevatron}},  {\em Comput. Phys.
  Commun.} {\bf 182} (2011) 2605--2631,
  [\href{http://arxiv.org/abs/1102.1898}{{\tt arXiv:1102.1898}}].

\bibitem{Bechtle:2013wla}
P.~Bechtle, O.~Brein, S.~Heinemeyer, O.~St\r{a}l, T.~Stefaniak, G.~Weiglein,
  and K.~E. Williams, {\it {$\mathsf{HiggsBounds}-4$: Improved Tests of
  Extended Higgs Sectors against Exclusion Bounds from LEP, the Tevatron and
  the LHC}},  {\em Eur. Phys. J. C} {\bf 74} (2014), no.~3 2693,
  [\href{http://arxiv.org/abs/1311.0055}{{\tt arXiv:1311.0055}}].

\bibitem{Bechtle:2020pkv}
P.~Bechtle, D.~Dercks, S.~Heinemeyer, T.~Klingl, T.~Stefaniak, G.~Weiglein, and
  J.~Wittbrodt, {\it {HiggsBounds-5: Testing Higgs Sectors in the LHC 13 TeV
  Era}},  {\em Eur. Phys. J. C} {\bf 80} (2020), no.~12 1211,
  [\href{http://arxiv.org/abs/2006.06007}{{\tt arXiv:2006.06007}}].

\bibitem{HFLAV:2022pwe}
{\bf HFLAV} Collaboration, Y.~S. Amhis et~al., {\it {Averages of $b$-hadron,
  $c$-hadron, and $\tau$-lepton properties as of 2021}},  {\em Phys. Rev. D}
  {\bf 107} (2023) 052008, [\href{http://arxiv.org/abs/2206.07501}{{\tt
  arXiv:2206.07501}}].

\bibitem{Hahn:2000kx}
T.~Hahn, {\it {Generating Feynman diagrams and amplitudes with FeynArts 3}},
  {\em Comput. Phys. Commun.} {\bf 140} (2001) 418--431,
  [\href{http://arxiv.org/abs/hep-ph/0012260}{{\tt hep-ph/0012260}}].

\bibitem{HAHN1999153}
T.~Hahn and M.~Pérez-Victoria, {\it Automated one-loop calculations in four
  and d dimensions},  {\em Computer Physics Communications} {\bf 118} (1999),
  no.~2 153--165.

\bibitem{vanOldenborgh:1990yc}
G.~J. van Oldenborgh, {\it {FF: A Package to evaluate one loop Feynman
  diagrams}},  {\em Comput. Phys. Commun.} {\bf 66} (1991) 1--15.

\bibitem{Hahn:2010zi}
T.~Hahn, {\it {Feynman Diagram Calculations with FeynArts, FormCalc, and
  LoopTools}},  {\em PoS} {\bf ACAT2010} (2010) 078,
  [\href{http://arxiv.org/abs/1006.2231}{{\tt arXiv:1006.2231}}].

\bibitem{ATLAS:2022tnm}
{\bf ATLAS} Collaboration, G.~Aad et~al., {\it {Measurement of the properties
  of Higgs boson production at $\sqrt{s} = 13$ TeV in the $H\to\gamma\gamma$
  channel using $139$ fb$^{-1}$ of $pp$ collision data with the ATLAS
  experiment}},  {\em JHEP} {\bf 07} (2023) 088,
  [\href{http://arxiv.org/abs/2207.00348}{{\tt arXiv:2207.00348}}].

\bibitem{ATLAS:2020qcv}
{\bf ATLAS} Collaboration, G.~Aad et~al., {\it {A search for the $Z\gamma$
  decay mode of the Higgs boson in $pp$ collisions at $\sqrt{s}$ = 13 TeV with
  the ATLAS detector}},  {\em Phys. Lett. B} {\bf 809} (2020) 135754,
  [\href{http://arxiv.org/abs/2005.05382}{{\tt arXiv:2005.05382}}].

\bibitem{CMS:2022ahq}
{\bf CMS} Collaboration, A.~Tumasyan et~al., {\it {Search for Higgs boson
  decays to a Z boson and a photon in proton-proton collisions at $ \sqrt{s} $
  = 13 TeV}},  {\em JHEP} {\bf 05} (2023) 233,
  [\href{http://arxiv.org/abs/2204.12945}{{\tt arXiv:2204.12945}}].

\bibitem{CMS:2023mku}
{\bf CMS, ATLAS} Collaboration, G.~Aad et~al., {\it {Evidence for the Higgs
  boson decay to a $Z$ boson and a photon at the LHC}},
  \href{http://arxiv.org/abs/2309.03501}{{\tt arXiv:2309.03501}}.

\bibitem{ATLAS:2023wqy}
{\bf ATLAS} Collaboration, G.~Aad et~al., {\it {Search for the $Z\gamma$ decay
  mode of new high-mass resonances in $pp$ collisions at $\sqrt{s}=13$ TeV with
  the ATLAS detector}},  \href{http://arxiv.org/abs/2309.04364}{{\tt
  arXiv:2309.04364}}.

\end{thebibliography}\endgroup
\end{document}